\documentclass[12pt]{amsart}
\usepackage{amsmath,amssymb,cite,enumerate,graphicx,hyperref}

\hypersetup{pdftex,colorlinks=true,allcolors=blue}
\usepackage[all]{hypcap}

\usepackage[foot]{amsaddr}
\usepackage[sc]{mathpazo}

\newcommand\ack{\subsection*{Acknowledgment}}

\DeclareMathAlphabet\mathsfbi{T1}{phv}{b}{it}

\numberwithin{equation}{section}

\newcommand\BV{\boldsymbol} % Vector bold
\newcommand\BM{\mathsfbi} % Matrix and tensor bold

\newcommand\dif{\:\!\mathrm{d}}
\newcommand\deriv[2]{\frac{\mathrm{d} #1}{\mathrm{d} #2}}
\newcommand\parderiv[2]{\frac{\partial #1}{\partial #2}}
\newcommand\coll{\mathcal C}
\newcommand\EE{\mathbb E}
\newcommand\RR{\mathbb R}
\newcommand\ZZ{\mathbb Z}
\newcommand\cL{\mathcal L}
\newcommand\cE{\mathcal E}
\newcommand\cF{\mathcal F}

\newcommand\PP{\mathbb P}

\newcommand\Hside{\:\!\Theta}

\newcommand\Mf{g}

\newcommand\myatop[2]{\genfrac{}{}{0pt}{}{#1}{#2}}

\begin{document}

\author[Rafail V. Abramov]{Rafail V. Abramov}

\address{Department of Mathematics, Statistics and Computer Science,
University of Illinois at Chicago, 851 S. Morgan st., Chicago, IL 60607}

\email{abramov@uic.edu}

\title{The random gas of hard spheres}

\begin{abstract}
The inconsistency between the time-reversible Liouville equation and
time-irreversible Boltzmann equation has been pointed out long ago by
Loschmidt. To avoid Loschmidt's objection, here we propose a new
dynamical system to model the motion of atoms of gas, with their
interactions triggered by a random point process. Despite being
random, this model can approximate the collision dynamics of rigid
spheres via adjustable parameters. We compute the exact statistical
steady state of the system, and determine the form of its marginal
distributions for a large number of spheres. We find that the
Kullback--Leibler entropy (a generalization of the conventional
Boltzmann entropy) of the full system of random gas spheres is a
nonincreasing function of time. Unlike the conventional hard sphere
model, the proposed random gas model results in a variant of the
Enskog equation, which is known to be a more accurate model of dense
gas than the Boltzmann equation. We examine the hydrodynamic limit of
the derived Enskog equation for spheres of constant mass density, and
find that the corresponding Enskog--Euler and Enskog--Navier--Stokes
equations acquire additional effects in both the advective and viscous
terms. In the dilute gas approximation, the Enskog equation simplifies
to the Boltzmann equation, while the Enskog--Euler and
Enskog--Navier--Stokes equations become the conventional Euler and
Navier--Stokes equations.
\end{abstract}

\maketitle

\section*{Introduction}

It is known that the atoms in an electrostatically neutral monatomic
gas interact via the Lennard-Jones potential \cite{Len}. At short
range, the Lennard-Jones potential is inversely proportional to the
12th power of the distance between the centers of two interacting
atoms. Due to this high order singularity, the mean field
approximation \cite{Vla}, which is often used to close the
Bogoliubov--Born--Green--Kirkwood--Yvon (BBGKY) hierarchy
\cite{Bog,BorGre,Kir} for the long-range (that is, electrostatic or
gravitational) potentials, cannot be applied to the molecular
collisions, since the relevant spatial integrals diverge as the
distance between the atoms approaches zero. To work around this issue,
in molecular kinetic theory the Lennard-Jones potential interaction is
replaced with the hard sphere collision model
\cite{ChaCow,Cer,Cer2,CerIllPul,Gra,GalRayTex}. According to this
model, the gas molecules are presumed to be rigid impenetrable
spheres, which interact instantaneously and elastically according to
the mechanics of the collision of two hypothetical billiard balls,
which have nonzero mass, but do not possess the moment of inertia.

In the conventional setting of the hard sphere gas dynamics, one
starts with the Liouville equation \cite{Cer,Cer2,CerIllPul,GalRayTex}
for the probability density function of the complete system of all
participating spheres. The Liouville equation is a homogeneous
transport equation, whose characteristics are the straight lines of
free flight of the spheres. In order to describe the collisions, the
Liouville equation is endowed with special deflection conditions at
the collision surface (the set of points in the coordinate space where
a pair of spheres is separated by their diameter). These deflection
conditions restrict the solutions of the Liouville equation to those
for which the probability of a pair of spheres entering the collision
surface equals the probability of a pair of spheres simultaneously
exiting the collision surface along the corresponding direction of
deflection. Assuming that the Liouville equation is already solved
with the solution being known, one then computes the corresponding
BBGKY hierarchy \cite{Bog,BorGre,Kir}. The lowest-order identity in
the BBGKY hierarchy is then converted to the Boltzmann equation
\cite{Bol,ChaCow,Cer,Cer2,CerIllPul,GalRayTex,Gra,Gols} via an
approximation known as the Boltzmann hierarchy, and the subsequent
factorization of the joint probability density of two spheres into the
product of single-sphere densities.

Upon the examination of the standard closure of hard sphere dynamics
\cite{Cer2,CerIllPul,GalRayTex}, we observe that the derivation of the
Boltzmann equation from the leading order BBGKY identity violates the
assumptions under which the BBGKY identity itself was derived. Also,
the observed contradictions are similar to those pointed out by
Loschmidt \cite{Los}. Yet, the Boltzmann equation is known in practice
to yield an accurate approximation to the observable gas dynamics.
Thus, we propose that the Boltzmann equation instead originates, in a
consistent manner, from a different model of hard sphere dynamics,
which, on one hand, does not violate Loschmidt's observation, but, on
the other hand, is on par with the hard sphere model at approximating
real molecular interactions.

To avoid Loschmidt's objection \cite{Los} and derive the Boltzmann
equation in a consistent fashion, we propose a new random process to
model the underlying hard sphere dynamics, where the changes in the
velocities of the spheres still obey the mechanics of the rigid sphere
collision, but the ``collisions'' themselves are triggered by a point
process. This random dynamical system possesses the infinitesimal
generator, so that the corresponding forward Kolmogorov equation for
its probability density is readily available via the integration by
parts. The proposed random hard sphere process can approximate the
deterministic collision process by increasing the intensity of the
triggering point process via a parameter. We compute some of the
steady states of the proposed process, and find that, while the
conventional Boltzmann entropy can both increase and decrease,
depending on a solution, the Kullback--Leibler entropy \cite{KulLei}
between a solution and the steady state is a nonincreasing function of
time. We also examine the structure of marginal distributions of the
steady state in the limit of infinitely many spheres.

Then, we compute the forward equation for a single-sphere marginal
distribution, under the assumption that the corresponding multisphere
probability density is invariant under an arbitrary reordering of
spheres. The closure that we use is, however, different from
conventional -- rather than using a direct factorization of the
probability state (used, for example, in the conventional transition
from the Boltzmann hierarchy to the Boltzmann equation), we take into
account the structure of the previously computed steady state of the
full multisphere system. We then find that, in the limit as the
intensity of the point process increases to infinity, a variant of the
Enskog equation \cite{BelLac,Ens,GapGer,Lac,Rez,vBeiErn} emerges,
which, of course, simplifies to the Boltzmann equation in the dilute
gas approximation. We then examine the hydrodynamic limit of the
Enskog equation for spheres of constant mass density, as it appears to
be physically plausible for atoms of the noble gases \cite{CleRaiRei}.
In this limit, we find that the resulting Enskog--Euler and
Enskog--Navier--Stokes equations acquire additional nonvanishing
terms, which are not present in the conventional gas dynamics
equations originating from the Boltzmann equation
\cite{Cer,Cer2,CerIllPul,Gra,Gols}. These additional effects disappear
in the dilute gas approximation, which yields the usual Boltzmann,
Euler, and Navier--Stokes equations, respectively.

\section{The hard sphere collision model and the Boltzmann equation}
\label{sec:conventional}

For the sake of clarity of the exposition, we start by citing the
standard derivation of the Boltzmann equation from the collision
mechanics of hard spheres \cite{Cer,Cer2,CerIllPul,GalRayTex}.  For
simplicity, we consider only two identical hard spheres, each of
diameter $\sigma$. We denote by $\BV x$ and $\BV v$ the coordinate of
the center and velocity of the first sphere, respectively, and by $\BV
y$ and $\BV w$ the corresponding coordinate of the center and velocity
of the second sphere. In the absence of contact, the spheres maintain
their constant velocities $\BV v$ and $\BV w$, while their respective
coordinates $\BV x$ and $\BV y$ are given via
\begin{equation}
\label{eq:xyt}
\deriv{\BV x}t=\BV v,\qquad\deriv{\BV y}t=\BV w.
\end{equation}
Whenever the distance $\|\BV x-\BV y\|$ between the
centers of the two spheres equals their diameter $\sigma$, their
velocities are changed, at that instance of time, to
\begin{subequations}
\label{eq:vwxy}
\begin{equation}
\BV v'=\BV v+(\BV w-\BV v)\cdot(\BV x-\BV y)\frac{\BV x-\BV y}{\|\BV
  x-\BV y\|^2},
\end{equation}
\begin{equation}
\BV w'=\BV w+(\BV v-\BV w)\cdot(\BV x-\BV y)\frac{\BV x-\BV y}{\|\BV
  x-\BV y\|^2},
\end{equation}
\end{subequations}
where $\BV v'$ and $\BV w'$ are the new values of velocities. Such a
transformation preserves the momentum and kinetic energy of the system
of the two spheres:
\begin{equation}
\label{eq:m_E}
\BV v'+\BV w'=\BV v+\BV w,\qquad \|\BV v'\|^2+\|\BV w'\|^2=\|\BV v\|^2
+\|\BV w\|^2.
\end{equation}
Here and below, we assume that the total momentum of the system (the
sum of velocities of the spheres) is zero without loss of generality,
as otherwise the momentum can be set to zero via a suitable Galilean
shift of the reference frame.

The transformation above in~\eqref{eq:vwxy} is fully symmetric;
indeed, subtracting the first relation from the second, we obtain
\begin{equation}
\BV v'-\BV w'=\BV v-\BV w+2(\BV w-\BV v)\cdot(\BV x-\BV y)\frac{\BV
  x-\BV y}{\|\BV x-\BV y\|^2}.
\end{equation}
Scalar-multiplying by $(\BV x-\BV y)/\|\BV x-\BV y\|$ on both sides,
we further obtain
\begin{equation}
\label{eq:vwxyn}
(\BV v'-\BV w')\cdot\frac{\BV x-\BV y}{\|\BV x-\BV y\|}=-(\BV v-\BV
w)\cdot\frac{\BV x-\BV y}{\|\BV x-\BV y\|}.
\end{equation}
Substituting the above expression into~\eqref{eq:vwxy}, we arrive at
\begin{subequations}
\label{eq:vwxy_inv}
\begin{equation}
\BV v=\BV v'+(\BV w'-\BV v')\cdot(\BV x-\BV y)\frac{\BV x-\BV y}{\|\BV
  x-\BV y\|^2},
\end{equation}
\begin{equation}
\BV w=\BV w'+(\BV v'-\BV w')\cdot(\BV x-\BV y)\frac{\BV x-\BV y}{\|\BV
  x-\BV y\|^2}.
\end{equation}
\end{subequations}
It is also easy to see that the Jacobian of the change of variables
$(\BV v,\BV w)\to (\BV v',\BV w')$ is unity; indeed, observe that, for
$(\BV x-\BV y)$ taken as a fixed parameter,
\begin{equation}
\label{eq:det1}
\det\left(\parderiv{(\BV v',\BV w')}{(\BV v,\BV w)}\right)=-1,
\end{equation}
which can be verified via the rank-one update lemma for determinants.

\subsection{The Liouville problem for two spheres}

Let $F(t,\BV x,\BV y,\BV v,\BV w)$ denote the probability density
function of the system of two spheres. Then, $F$ satisfies the
following conditions:
\begin{itemize}
\item If the distance between the spheres is less than their diameter
  (that is, the spheres are overlapped), then the density is set to
  zero:
\begin{equation}
\label{eq:F_zero}
F(\BV x,\BV y,\BV v,\BV w)=0,\qquad\|\BV x-\BV y\|<\sigma.
\end{equation}
This condition ascertains that the spheres are rigid and may not
overlap.
\item If the distance between the spheres is greater than their
  diameter (that is, the spheres are separated), then $F$ obeys the
  Liouville equation \cite{Cer,Cer2,CerIllPul}:
\begin{equation}
\label{eq:liouville}
\parderiv Ft+\BV v\cdot\parderiv F{\BV x}+\BV w\cdot\parderiv F{\BV
  y}=0,\qquad\|\BV x-\BV y\|>\sigma.
\end{equation}
\item The Liouville equation \eqref{eq:liouville} is solved in the
  open set $\|\BV x-\BV y\|>\sigma$ -- therefore, a boundary condition
  is needed at the collision surface $\|\BV x-\BV y\|=\sigma$. This
  boundary condition is given via
\begin{equation}
\label{eq:BC}
F(\BV x,\BV y,\BV v',\BV w')=F(\BV x,\BV y,\BV v,\BV w),\qquad\|\BV
x-\BV y\|=\sigma,
\end{equation}
where $\BV v'$ and $\BV w'$ are given in \eqref{eq:vwxy} as functions
of $\BV x$, $\BV y$, $\BV v$ and $\BV w$.
\item Observe that in the open set $\|\BV x-\BV y\|<\sigma$, $F=0$
  also solves the Liouville equation in \eqref{eq:liouville} -- albeit
  with zero initial and boundary conditions. This leads to a possible
  discontinuity of $F$ at the collision surface $\|\BV x-\BV
  y\|=\sigma$, which will be taken into account below.
\item Assuming that there is a solution of \eqref{eq:liouville} in
  both open regions $\|\BV x-\BV y\|<\sigma$ and $\|\BV x-\BV
  y\|>\sigma$ (with boundary conditions given via zero and
  \eqref{eq:BC}, respectively), on the collision surface $\|\BV x-\BV
  y\|=\sigma$ itself we assign $F$ to be equal to the ``outer''
  boundary condition \eqref{eq:BC}. This definition of $F$ at the
  collision surface means that $F$ is continuous as $\|\BV x-\BV
  y\|\downarrow\sigma$, and possibly discontinuous as $\|\BV x-\BV
  y\|\uparrow\sigma$.
\end{itemize}
In what follows, we assume for convenience that the spatial part of
the domain has a finite volume, but no boundary effects other than
those in \eqref{eq:BC}. For example, one can assume that the space of
coordinates is periodic (that is, if a sphere leaves the coordinate
``box'' through a wall, it immediately re-enters from the opposite
wall).

According to \eqref{eq:BC}, the probability that a pair of spheres
enters the collision surface at the point $(\BV x,\BV y)$ and time $t$
with velocities $(\BV v,\BV w)$ is the same as that of a pair of
spheres exiting at the same point and time, but with the velocities
$(\BV v',\BV w')$.  The condition in \eqref{eq:BC} preserves the
normalization of $F$ separately in the region $\|\BV x-\BV
y\|>\sigma$.

Note that the formulation of the problem in
\eqref{eq:F_zero}--\eqref{eq:BC} is not fully equivalent to the actual
dynamics in \eqref{eq:xyt} and \eqref{eq:vwxy}. For the formulation
\eqref{eq:F_zero}--\eqref{eq:BC} to correspond precisely to
\eqref{eq:xyt} and \eqref{eq:vwxy}, one needs the characteristic
curves of the Liouville equation in \eqref{eq:liouville} to be the
trajectories of the system in \eqref{eq:xyt} and \eqref{eq:vwxy}.
However, this is not the case in \eqref{eq:F_zero}--\eqref{eq:BC} --
the characteristic curves are straight lines given via constant
parameters $\BV v$ and $\BV w$, which pierce the collision surface
$\|\BV x-\BV y\|=\sigma$.  Instead, the ``collisions'' in
\eqref{eq:F_zero}--\eqref{eq:BC} are implemented via reassigning the
values of $F$ between different characteristics according to
\eqref{eq:BC}.

\subsection{The BBGKY identity for two spheres}

Here we follow \cite{CerIllPul} and derive the BBGKY hierarchy (which,
for two spheres, consists of a single identity) and, subsequently, the
Boltzmann equation. In what follows, we take advantage of the fact
that $F=0$ inside the overlapped region $\|\BV x-\BV y\|<\sigma$ also
satisfies the Liouville equation in \eqref{eq:liouville} with zero
initial and boundary conditions.

Let us assume that a solution $F$ is computed for the Liouville
problem \eqref{eq:F_zero}--\eqref{eq:BC} for both overlapped and
non-overlapped regions, so that the relation in \eqref{eq:liouville}
is no longer an equation, but rather an identity. In what follows, we
assume that $F$ is symmetric under the permutations of two spheres --
that is, $F(t,\BV x,\BV y,\BV v,\BV w)=F(t,\BV y,\BV x,\BV w,\BV
v)$. Our goal is to manipulate the Liouville identity in
\eqref{eq:liouville} so as to obtain an appropriate identity for the
marginal distribution of a single sphere
\begin{equation}
\label{eq:f_marginal}
f(t,\BV x,\BV v)=\int F(t,\BV x,\BV y,\BV v,\BV w)\dif\BV y\dif\BV w.
\end{equation}
First, we integrate the identity in \eqref{eq:liouville} in $\dif\BV
y\dif\BV w$ over the whole space, which includes both overlapped (that
is, $\|\BV x-\BV y\|<\sigma$) and non-overlapped (that is, $\|\BV
x-\BV y\|>\sigma$) regions:
\begin{equation}
\label{eq:liouville_dydw}
\int\left(\parderiv Ft+\BV v\cdot\parderiv F{\BV x}+\BV w\cdot
\parderiv F{\BV y}\right)\dif\BV y\dif\BV w=0.
\end{equation}
Next, we exchange the order of differentiation and integration
above. For the $t$-derivative term, the exchange is done in a direct
manner, since the $\dif\BV y\dif\BV w$-integration is unrelated to
$t$:
\begin{equation}
\label{eq:t_deriv}
\int\parderiv Ft\dif\BV y\dif\BV w=\parderiv{}t\int F\dif\BV y\dif\BV
w=\parderiv ft.
\end{equation}
However, due to the discontinuity in $F$ at the collision surface
$\|\BV x-\BV y\|=\sigma$, one cannot directly swap the spatial
derivatives and the spatial integration. For the term with $\BV
y$-derivative, we use the Gauss theorem:
\begin{multline}
\label{eq:y_deriv}
\int\BV w\cdot\parderiv F{\BV y}\dif\BV y\dif\BV w=\int\dif\BV w
\left(\int_{\|\BV x-\BV y\|<\sigma}\parderiv{}{\BV y}\cdot(F\BV w)\dif
\BV y+\int_{\|\BV x-\BV y\|>\sigma}\parderiv{}{\BV y}\cdot(F\BV w)\dif
\BV y\right)=\\=\int\dif\BV w\int_{\|\BV x-\BV y\|=\sigma}(
F_{\eqref{eq:F_zero}}-F_{\eqref{eq:BC}})\BV w\cdot\BV n_{\BV x}(\BV y)
\dif S_{\BV x}(\BV y)=-\int\dif\BV w\int_{\|\BV x-\BV y\|=\sigma}F\BV
w\cdot\BV n_{\BV x}(\BV y) \dif S_{\BV x}(\BV y).
\end{multline}
Above, $\dif S_{\BV x}(\BV y)$ is the surface area element at the
point $\BV y$ of the sphere of radius $\sigma$ centered at $\BV x$,
the unit normal vector $\BV n_{\BV x}(\BV y)$, which originates at
$\BV x$ in the direction of $\BV y$, is given via
\begin{equation}
\BV n_{\BV x}(\BV y)=\frac{\BV y-\BV x}{\|\BV y-\BV x\|}.
\end{equation}
Also, $F_{\eqref{eq:F_zero}}$ denotes the boundary value for $\|\BV
x-\BV y\|<\sigma$ (which is zero), $F_{\eqref{eq:BC}}$ denotes the
boundary value for $\|\BV x-\BV y\|>\sigma$, given via \eqref{eq:BC},
and we replace $F_{\eqref{eq:BC}}$ with $F$ in the last identity, as
it was agreed above that $F$ at the collision surface $\|\BV x-\BV
y\|=\sigma$ is assigned the boundary value given via \eqref{eq:BC}.

For the term with the $\BV x$-derivative, we need to use the
generalized Leibniz rule. First, we split
\begin{equation}
\int\BV v\cdot\parderiv F{\BV x}\dif\BV y\dif\BV w=\int\dif\BV w
\left(\int_{\|\BV x-\BV y\|>\sigma}\BV v\cdot\parderiv F{\BV x}\dif
\BV y+\int_{\|\BV x-\BV y\|<\sigma}\BV v\cdot\parderiv F{\BV x}\dif
\BV y\right).
\end{equation}
Then, for the first term in the right-hand side we write
\begin{multline}
\int_{\|\BV x-\BV y\|>\sigma}\BV v\cdot\parderiv{}{\BV x}F(\BV x,\BV
y)\dif\BV y=\int_{\|\BV z\|>\sigma}\BV v\cdot\left(\parderiv{}{\BV
  x}-\parderiv{}{\BV z}\right)F(\BV x,\BV x+\BV z)\dif\BV z=\\=\BV
v\cdot\parderiv{}{\BV x}\int_{\|\BV z\|>\sigma}F(\BV x,\BV x+\BV
z)\dif\BV z-\int_{\|\BV z\|>\sigma}\parderiv{}{\BV z}\cdot(F(\BV x,\BV
x+\BV z)\BV v)\dif\BV z=\\=\BV v\cdot\parderiv{}{\BV x}\int_{\|\BV x-
  \BV y\|>\sigma}F(\BV x,\BV y)\dif\BV y+\int_{\|\BV x-\BV y\|=\sigma}
F_{\eqref{eq:BC}}\BV v\cdot\BV n_{\BV x}(\BV y) \dif S_{\BV x}(\BV y).
\end{multline}
For the second term, we repeat the calculations in a similar way:
\begin{equation}
\int_{\|\BV x-\BV y\|<\sigma}\BV v\cdot\parderiv{}{\BV x}F(\BV x,\BV
y)\dif\BV y=\BV v\cdot\parderiv{}{\BV x}\int_{\|\BV x-\BV y\|<\sigma}
F(\BV x,\BV y)\dif\BV y-\int_{\|\BV x-\BV y\|=\sigma} F_{
  \eqref{eq:F_zero}}\BV v\cdot\BV n_{\BV x}(\BV y)\dif S_{\BV x}(\BV
y).
\end{equation}
Adding the two terms, recalling that $F_{\eqref{eq:F_zero}}$ is zero,
$F_{\eqref{eq:BC}}$ is $F$ itself on $\|\BV x-\BV y\|=\sigma$, and
using \eqref{eq:f_marginal}, we arrive at the identity
\begin{equation}
\int\BV v\cdot\parderiv F{\BV x}\dif\BV y\dif\BV w=\BV v\cdot\parderiv
f{\BV x}+\int\dif\BV w\int_{\|\BV x-\BV y\|=\sigma}F\BV v\cdot\BV
n_{\BV x}(\BV y)\dif S_{\BV x} (\BV y).
\end{equation}
Combining the above expression with \eqref{eq:t_deriv} and
\eqref{eq:y_deriv}, and substituting into \eqref{eq:liouville_dydw},
we arrive at the identity
\begin{equation}
\parderiv ft+\BV v\cdot\parderiv f{\BV x}=\int\dif\BV w\int_{\|\BV x-
  \BV y\|=\sigma}F(\BV w-\BV v)\cdot\BV n_{\BV x}(\BV y)\dif S_{\BV x}
(\BV y).
\end{equation}
For further convenience, we can rename the dummy variables of
integration so that the surface integration is computed over a unit
sphere (that is, over $\dif\BV n$). Namely, since $\BV y$ follows the
surface of the sphere of radius $\sigma$ centered at $\BV x$, we
denote
\begin{equation}
\BV y=\BV x+\sigma\BV n,\qquad\dif S_{\BV x}(\BV y)=\sigma^2\dif\BV n.
\end{equation}
In the above variables, the identity in \eqref{eq:vwxy} becomes
\begin{equation}
\label{eq:vwn}
\BV v'=\BV v+\big((\BV w-\BV v)\cdot\BV n\big)\BV n,\qquad \BV w'=\BV
w+\big((\BV v-\BV w)\cdot\BV n\big)\BV n,
\end{equation}
where we note that there is no dependence on $\sigma$. The integral
over the collision surface, written in the new variables, yields the
following identity:
\begin{equation}
\label{eq:BBGKY-1}
\parderiv ft+\BV v\cdot\parderiv f{\BV x}=\sigma^2\int F(t,\BV x, \BV
x+\sigma\BV n,\BV v,\BV w)(\BV w-\BV v)\cdot\BV n\dif\BV n\dif \BV w.
\end{equation}
In order to transform the surface integral in the right-hand side
above into the Boltzmann collision integral, one further needs to
rearrange the surface integration with the help of \eqref{eq:BC}.
Consider the distance $\|\BV y(t)-\BV x(t)\|$ between the spheres as a
function of time. The sign of its time derivative indicates whether
the spheres are approaching or escaping each other:
\begin{equation}
\label{eq:dxyt}
\deriv{}t\|\BV y(t)-\BV x(t)\|=\frac{\BV y-\BV x}{\|\BV y-\BV
  x\|}\cdot(\BV w-\BV v)=(\BV w-\BV v)\cdot\BV n.
\end{equation}
Clearly, whenever the time derivative above is positive (that is, the
spheres escape each other), the dot-product $(\BV w-\BV v)\cdot\BV n$
is positive, and vice versa. Following
\cite{Cer2,CerIllPul,GalRayTex}, we use the condition in \eqref{eq:BC}
for $F$ and rearrange the collision integral in \eqref{eq:BBGKY-1},
with help of the Heaviside step-function $\Hside(x)$, as follows:
\begin{multline}
\sigma^2\int F(t,\BV x,\BV x+\sigma\BV n,\BV v,\BV w)(\BV w-\BV v)
\cdot\BV n\dif\BV n\dif \BV w=\\=\sigma^2\int F(t,\BV x,\BV x+\sigma
\BV n,\BV v',\BV w')(\BV w-\BV v)\cdot\BV n\Hside\big((\BV w-\BV v)
\cdot\BV n\big)\dif\BV n\dif \BV w+\\+\sigma^2\int F(t,\BV x,\BV x+
\sigma\BV n,\BV v,\BV w)(\BV w-\BV v)\cdot\BV n\Hside\big((\BV v-\BV
w)\cdot\BV n\big)\dif\BV n\dif\BV w=\\=\sigma^2\int\big(F(t,\BV x,\BV
x+\sigma\BV n,\BV v',\BV w')-F(t,\BV x,\BV x-\sigma\BV n,\BV v,\BV w)
\big)(\BV w-\BV v)\cdot\BV n\Hside\big((\BV w-\BV v)\cdot\BV n\big)
\dif\BV n\dif\BV w.
\end{multline}
Above, we split the collision integral into two hemispheres -- one
where $(\BV w-\BV v)\cdot\BV n>0$ (that is, the spheres are escaping
each other), and another one where the spheres are approaching each
other. In the first hemisphere, we replace $F(\BV v,\BV w)$ with
$F(\BV v',\BV w')$, with $\BV v'=\BV v'(\BV v,\BV w,\BV n)$ and $\BV
w'=\BV w'(\BV v,\BV w,\BV n)$ given via \eqref{eq:vwn}. This is a
valid rearrangement since the condition \eqref{eq:BC} requires $F(\BV
v,\BV w)$ and $F(\BV v',\BV w')$ to be equal anywhere on the surface
of integration. In the second hemisphere, we change the sign of $\BV
n$ to the opposite.  Then, we merge the integrals. The resulting BBGKY
identity is given via
\begin{multline}
\label{eq:BBGKY2}
\parderiv ft+\BV v\cdot\parderiv f{\BV x}=\sigma^2\int\big(F(t,\BV x,
\BV x+\sigma\BV n,\BV v',\BV w')-\\-F(t,\BV x,\BV x-\sigma\BV n,\BV v,
\BV w)\big)(\BV w-\BV v)\cdot\BV n\Hside\big((\BV w-\BV v)\cdot\BV n
\big)\dif\BV n\dif\BV w.
\end{multline}
Clearly, the BBGKY identity above is the same identity as in
\eqref{eq:BBGKY-1}, albeit rewritten in the form
\eqref{eq:BBGKY2}. The transition from \eqref{eq:BBGKY-1} to
\eqref{eq:BBGKY2} is justified for any $F$ which satisfies
\eqref{eq:BC} on the collision surface, with $f$ being the marginal
distribution of $F$ given via \eqref{eq:f_marginal}.

\subsection{The Boltzmann hierarchy and Boltzmann equation}

The derivation above can be extended to $K$ spheres
\cite{Cer2,CerIllPul,GalRayTex}, assuming that $F$ is symmetric under
an arbitrary permutation of the spheres. This results in the multiple
BBGKY identities for marginal distributions of various orders,
chain-linked to each other starting with the highest-order. The
lowest-order identity in the BBGKY hierarchy for $K$ spheres is given
via
\begin{multline}
\label{eq:BBGKYK}
\parderiv ft+\BV v\cdot\parderiv f{\BV x}=(K-1)\sigma^2\int\big(
F^{(2)}(t, \BV x, \BV x+\sigma\BV n,\BV v',\BV w')-\\-F^{(2)}(t,\BV
x,\BV x-\sigma\BV n,\BV v,\BV w)\big)(\BV w-\BV v)\cdot\BV n\Hside
\big((\BV w-\BV v) \cdot\BV n \big)\dif\BV n\dif\BV w,
\end{multline}
where $F^{(2)}$ is the two-sphere marginal distribution of the full
$K$-sphere density $F$:
\begin{equation}
\label{eq:F2_marginal}
F^{(2)}(t,\BV x_1,\BV x_2,\BV v_1,\BV v_2)=\int F(t,\BV x_1,\ldots\BV
x_K,\BV v_1,\ldots,\BV v_K)\dif\BV x_3\dif\BV x_K\ldots\dif\BV v_3\dif
\BV v_K.
\end{equation}
The Boltzmann equation \cite{Bol,Cer,Cer2,CerIllPul,Gra,Gols} is
obtained from the BBGKY identity \eqref{eq:BBGKYK} in two following
steps. First, one assumes that
\begin{equation}
\label{eq:boltzmann-grad}
\sigma\to 0,\qquad K\to\infty,\qquad K\sigma^2\to\text{constant},
\end{equation}
which is known as the Boltzmann--Grad limit \cite{Gra}. As $\sigma\to
0$, the following assumption is made \cite{CerIllPul,GalRayTex}:
\begin{equation}
\label{eq:step1}
F^{(2)}(t,\BV x,\BV x- \sigma\BV n,\BV v,\BV w)\approx F^{(2)}(t,\BV x,\BV
x,\BV v,\BV w),\quad F^{(2)}(t,\BV x,\BV x+\sigma\BV n,\BV v',\BV
w')\approx F^{(2)}(t,\BV x,\BV x,\BV v',\BV w').
\end{equation}
This assumption transforms the BBGKY hierarchy into what is known as
the Boltzmann hierarchy \cite{CerIllPul,GalRayTex}, whose lowest-order
relation is given via
\begin{multline}
\label{eq:boltzmann_hierarchy}
\parderiv ft+\BV v\cdot\parderiv f{\BV x}=(K-1)\sigma^2\int\big(
F^{(2)}(t,\BV x,\BV x,\BV v',\BV w')-\\-F^{(2)}(t,\BV x,\BV x,\BV
v,\BV w)\big) (\BV w-\BV v)\cdot\BV n\Hside\big((\BV w-\BV v)\cdot\BV
n\big)\dif\BV n\dif\BV w.
\end{multline}
Next, the joint two-sphere marginal density $F^{(2)}$ above is
approximated as follows:
\begin{equation}
\label{eq:step2}
F^{(2)}(t,\BV x,\BV x,\BV v,\BV w)=f(t,\BV x,\BV v)f(t,\BV x,\BV w),
\quad F^{(2)}(t,\BV x,\BV x,\BV v',\BV w')=f(t,\BV x,\BV v')f(t,\BV x,
\BV w').
\end{equation}
Substituting the approximations in \eqref{eq:step2} into
\eqref{eq:boltzmann_hierarchy}, one arrives at
\begin{multline}
\label{eq:boltzmann}
\parderiv ft+\BV v\cdot\parderiv f{\BV x}=(K-1)\sigma^2\int\big(f(\BV
x,\BV v')f(\BV x,\BV w')-\\-f(\BV x,\BV v)f(\BV x,\BV w)\big)(\BV w-
\BV v)\cdot\BV n\Hside\big((\BV w-\BV v)\cdot\BV n\big)\dif\BV n\dif
\BV w.
\end{multline}
The relation above is known as the Boltzmann equation
\cite{Bol,Cer,Cer2,CerIllPul,Gols,Gra}. Unlike the BBGKY identity in
\eqref{eq:BBGKYK}, built upon a (presumably known) solution of the
Liouville problem for $K$ spheres, \eqref{eq:boltzmann} is treated as
a closed, self-contained equation for the marginal distribution $f$.
The factor $(K-1)\sigma^2$ in front of the integral above can be
changed to $K\sigma^2$, since the Boltzmann equation is the result of
the Boltzmann--Grad limit in \eqref{eq:boltzmann-grad}.

\section{Inconsistencies in the derivation of the Boltzmann equation}

Here we point out some contradictions in the conventional derivation
of the Boltzmann equation \eqref{eq:boltzmann} from the Liouville
problem \eqref{eq:F_zero}--\eqref{eq:BC}, presented in the previous
section.  We start with a brief summary of the derivation.

\begin{enumerate}[\indent (1)]
\item The Liouville equation \eqref{eq:liouville} by itself does not
  have any collision effects. The effect of collision is imposed
  separately for every pair of spheres with coordinates $\BV x$ and
  $\BV y$, on the surface $\|\BV x-\BV y\|=\sigma$.
\item Due to the effect on the collision surface, the probability
  density of states $F$ is discontinuous on this surface -- it is zero
  for all $\|\BV x-\BV y\|<\sigma$ according to \eqref{eq:F_zero} (as
  the spheres are impenetrable) and is generally nonzero otherwise.
\item Due to the discontinuity of $F$ on the collision surface, the
  collision surface integrals emerge according to the Gauss theorem
  and the Leibniz rule, when the BBGKY hierarchy is constructed. The
  resulting identity is given by \eqref{eq:BBGKY-1} -- observe that it
  is valid for any $F$, which satisfies \eqref{eq:F_zero},
  \eqref{eq:liouville}, and is discontinuous on the collision surface
  $\|\BV x-\BV y\|=\sigma$.
\item In addition to the discontinuity, the velocity deflection
  condition in \eqref{eq:BC} is imposed on $F$. This condition allows
  to rewrite the collision integral in \eqref{eq:BBGKY-1} in the form
  \eqref{eq:BBGKY2}. Note that it is the same exact integral as
  obtained originally in \eqref{eq:BBGKY-1} via Gauss theorem and
  Leibniz rule, only written in a different manner thanks to
  \eqref{eq:BC}. Similarly, one can obtain the $K$-sphere BBGKY
  identity in \eqref{eq:BBGKYK} \cite{CerIllPul,GalRayTex}.
\end{enumerate}
After the BBGKY identity is obtained in \eqref{eq:BBGKY2} (or its
$K$-sphere version \eqref{eq:BBGKYK}), the following additional steps
are taken to obtain the Boltzmann equation:
\begin{enumerate}[\indent (a)]
\item The integrand in the collision integral of \eqref{eq:BBGKYK} is
  modified via \eqref{eq:step1}, where the distance between the
  centers of colliding spheres is reduced to zero. The result is
  referred to as the Boltzmann hierarchy
  \eqref{eq:boltzmann_hierarchy}.
\item The already modified integrand in the collision integral of
  \eqref{eq:boltzmann_hierarchy} is modified again via
  \eqref{eq:step2}, where the two-sphere marginal distribution is
  replaced with the product of two single-sphere marginal
  distributions. This results in the Boltzmann equation
  \eqref{eq:boltzmann}.
\end{enumerate}
Remarkably, none of the steps in (a)--(b) above are consistent with
the conditions under which steps in (1)--(4) were valid. Below we
elaborate on the inconsistencies between (a)--(b) and the conditions
under which (1)--(4) were obtained.

\subsection{A contradiction between the Liouville problem and the
  Boltzmann hierarchy}

Observe that if $F$ is a distribution of $K$ rigid spheres, then
\begin{equation}
\label{eq:contradiction}
F^{(2)}(t,\BV x,\BV x,\BV v,\BV w)=0,\quad\text{for all }\BV x,\BV
v,\BV w\text{ and for any }\sigma.
\end{equation}
The reason for this is the following. Observe that $F^{(2)}$ with the
arguments given above is the two-sphere marginal distribution of the
full probability density $F$, where the latter is computed at the
point where the coordinates of the first and second spheres are
identical. In such a state, the first and second spheres are fully
overlapped for any value of the diameter $\sigma$. Due to the
impenetrability requirement in \eqref{eq:F_zero} (which can be
extended onto $K$ spheres in a direct manner), the value of $F$ at
this state is guaranteed to be zero irrespectively of the states of
all other spheres:
\begin{equation}
F(t,\BV x,\BV x,\BV x_3,\ldots,\BV x_K,\BV v_1,\ldots,\BV v_K)=0,\quad
\text{for all }\BV x,\BV x_3,\ldots,\BV x_K,\BV v_1,\ldots,\BV v_K.
\end{equation}
The identity above, in turn, means that the marginal integral
\eqref{eq:F2_marginal} of such an $F$ is also zero, which leads to
\eqref{eq:contradiction}.

Now, observe that the marginal distribution $F^{(2)}$, computed for
the states of the form in \eqref{eq:contradiction}, is used in the
collision integral of the lowest-order equation of the Boltzmann
hierarchy in \eqref{eq:boltzmann_hierarchy}, which was obtained from
the BBGKY equation in \eqref{eq:BBGKYK} via \eqref{eq:step1}.
Comparing the right-hand side of \eqref{eq:boltzmann_hierarchy} with
\eqref{eq:contradiction}, we find that the collision integral in
\eqref{eq:boltzmann_hierarchy} must always be zero, if the
impenetrability condition in \eqref{eq:F_zero} is indeed satisfied.

Therefore, if the approximation of the BBGKY equation
\eqref{eq:BBGKYK} via the Boltzmann hierarchy equation
\eqref{eq:boltzmann_hierarchy} is formally valid, this could only mean
that any physical effects from the collision integral of the BBGKY
equation in \eqref{eq:BBGKYK} somehow vanish in the Boltzmann--Grad
limit \eqref{eq:boltzmann-grad}. However, even if that is indeed the
case, it is not the dynamical regime which is physically interesting
or relevant -- in typical applications of gas dynamics, the molecular
collisions and the associated effects (such as viscosity and heat
conductivity) are usually important.

\subsection{A contradiction between the Liouville problem and the
Boltzmann closure}

Let us assume that the inconsistency between the Liouville problem in
\eqref{eq:F_zero}--\eqref{eq:BC} and the Boltzmann hierarchy in
\eqref{eq:boltzmann_hierarchy}, described above, can somehow be
``overlooked''. Despite that, yet another contradiction arises when
the integrand under the collision integral in
\eqref{eq:boltzmann_hierarchy} is replaced with the product of the
single-sphere marginal distributions via \eqref{eq:step2}. Recall that
the identities in \eqref{eq:BBGKY-1} and \eqref{eq:BBGKY2} are the
same exact identity, written in two different forms due to the fact
the integrand in the collision integral obeys \eqref{eq:BC} -- in
fact, the collision integral in \eqref{eq:BBGKY2} is the same exact
integral as in \eqref{eq:BBGKY-1}, rewritten in a different manner
thanks to \eqref{eq:BC}.

However, observe that the factorization in \eqref{eq:step2} does not
generally obey the conditions for which the rearrangement of the
collision integrals between \eqref{eq:BBGKY-1} and \eqref{eq:BBGKY2}
is valid. Indeed, if the integrands of the collision integrals in
\eqref{eq:BBGKY-1} and \eqref{eq:BBGKY2} are replaced with the
factorization in \eqref{eq:step2}, for the same transition to be valid
one must have
\begin{equation}
\label{eq:fvfw}
f(\BV x,\BV v)f(\BV x,\BV w)=f(\BV x,\BV v')f(\BV x,\BV w'), \quad
\text{for all }\BV x,\BV v,\BV w,\text{ and all unit vectors }\BV n,
\end{equation}
where $\BV v'=\BV v'(\BV v,\BV w,\BV n)$ and $\BV w'=\BV w'(\BV v,\BV
w,\BV n)$ are given via \eqref{eq:vwn}. However, solutions of the
Boltzmann equation in \eqref{eq:boltzmann} which are not simple
traveling waves clearly may not have such a property, since
\eqref{eq:fvfw} automatically turns the Boltzmann collision integral
in the right-hand side of \eqref{eq:boltzmann} into zero.

Additionally, observe that, since \eqref{eq:fvfw} automatically sets
the Boltzmann collision integral in \eqref{eq:boltzmann} to zero, any
solution, which is not a traveling wave, is {\em required} to violate
\eqref{eq:BC}. In other words, not only the violation of the
deflection condition in \eqref{eq:BC} is ``overlooked'' in the
Boltzmann equation, but it is, in fact, used as the quintessential
device to create its nontrivial solutions. This happens despite the
fact that \eqref{eq:BC} is the fundamental property of all solutions
of the Liouville problem, whether stationary or not, which, together
with \eqref{eq:F_zero}, leads to the BBGKY identity in
\eqref{eq:BBGKY2}, from which the Boltzmann collision integral and the
Boltzmann equation are subsequently derived.

Therefore, we conclude that what we observe in the conventional
derivation of the Boltzmann equation is a logical fallacy -- first, a
chain of identities is established under the assumption that certain
conditions hold; then, the resultant identity is replaced with an
equation whose nontrivial solutions {\em must} violate requisite
conditions under which the identity was derived to begin with.

\subsection{Reversibility and Loschmidt's objection}

One may try to ``escape'' the above reasoning by assuming that the
transitions from BBGKY to the Boltzmann hierarchy in \eqref{eq:step1}
and further to the factorization in \eqref{eq:step2} are only valid
for incident velocities, but not recedent. More precisely, one may put
forth an ansatz that, for recedent directions given via $\BV
n\cdot(\BV v-\BV w)>0$,
\begin{equation}
\label{eq:irreversibility_ansatz}
F^{(2)}(t,\BV x,\BV x-\sigma\BV n,\BV v,\BV w)\neq F^{(2)}(t,\BV x,\BV
x,\BV v,\BV w),\quad F^{(2)}(t,\BV x,\BV x+\sigma\BV n,\BV v',\BV w')
\neq F^{(2)}(t,\BV x,\BV x,\BV v',\BV w'),
\end{equation}
where $\BV v'$ and $\BV w'$ are functions of $\BV v$, $\BV w$ and $\BV
n$, given in \eqref{eq:vwn}. In such a case, one can claim that the
factorization in \eqref{eq:step2} ``is not required'' to satisfy
\eqref{eq:BC}, so that the relation \eqref{eq:fvfw} does not have to
apply between the incident and recedent directions.

First, we observe that the ansatz in \eqref{eq:irreversibility_ansatz}
is undoubtedly correct, and not just for the recedent, but also for
incident directions, since $F^{(2)}(t,\BV x,\BV x,\BV v,\BV w)=0$ for
all $\BV x$, $\BV v$, $\BV w$ and $\sigma$, as we already pointed out
above in \eqref{eq:contradiction}.  However, even if one somehow
``overlooks'' this fact, still such a hypothetical dichotomy between
the incident and recedent directions is not only unfounded, but is
also clearly in contradiction with the original formulation of the
Liouville problem \eqref{eq:F_zero}--\eqref{eq:BC}. The reason is that
in the Liouville problem all collisions are exactly reversible, and
all states of the system reproduce themselves in reverse order if one
uses the terminal coordinates and negatives of terminal velocities in
place of an initial condition.

In the case of such a reversal, the incident density states become
recedent and vice versa, however, the formulation of the Liouville
problem in \eqref{eq:F_zero}--\eqref{eq:BC} remains invariant.  Thus,
the same exact reasoning as above involving the BBGKY hierarchy, the
Boltzmann hierarchy in \eqref{eq:step1} and factorization
\eqref{eq:step2} can subsequently be applied to the reversed states as
well, in which case the formerly recedent states will be expressed via
\eqref{eq:step1} and \eqref{eq:step2}. Or, conversely, if the ansatz
in \eqref{eq:irreversibility_ansatz} indeed holds for recedent states,
then it also automatically holds for the incident states, in which
case the transition from \eqref{eq:BBGKYK} to \eqref{eq:boltzmann}
cannot be carried out via \eqref{eq:step1} and \eqref{eq:step2} (which
is in fact the case due to \eqref{eq:contradiction}).

A similar objection to the form of the Boltzmann collision integral
has been posed by Loschmidt \cite{Los}, who pointed out that, due to
the entropy inequality (also known as the $H$-theorem \cite{Bol}), the
Boltzmann equation is time-irreversible, while the underlying dynamics
of hard spheres, from which the collision integral is derived, are
time-reversible.  It is commonly known as {\em Loschmidt's paradox},
even though Loschmidt's observation is entirely valid and does not
constitute a ``paradox'' by itself.  Instead, the ``paradox'' appears
due to the logical fallacy hidden between the formulation of the
Liouville problem for hard spheres in \eqref{eq:F_zero}--\eqref{eq:BC}
and the resultant Boltzmann collision integral in
\eqref{eq:boltzmann}, as we exposed above.

\subsection{Our proposal to amend the situation}

At the same time, there is no doubt that the Boltzmann equation
\eqref{eq:boltzmann} is a very accurate model of a dilute gas, which
is confirmed by numerous observations and experiments. This means that
the collision integral in the right-hand side of \eqref{eq:boltzmann}
is a valid approximation of the statistical effects of molecular
interactions in practical scenarios, even if there is a conceptual
flaw in its conventional derivation \cite{Cer2,CerIllPul,GalRayTex}.
In what follows, we propose a different model of hard sphere
interactions in which such a problem does not manifest, and which also
leads to the Boltzmann equation in the dilute gas approximation --
albeit in a different way, via the Enskog equation
\cite{BelLac,Ens,GapGer,Lac,Rez,vBeiErn}.

From what is presented above, it is clear that the formulation of the
dynamics of the spheres needs to be such that the collision integrals
in the BBGKY equation \eqref{eq:BBGKYK} possess their form
independently of what their integrand is. For that, we need to
disentangle the instantaneous changes of sphere velocities from the
properties of $F$ in the Liouville equation, and arrange for them to
be the property of the equation itself. This is possible to do if the
underlying dynamical process possesses an {\em infinitesimal
  generator}.  Generally, let $\BV z(t)$ be a Markov process in a
$d$-dimensional Euclidean space. Assuming that $\BV z(t)=\BV z$ is
known at the time $t$, let us consider the conditional expectation
$\EE[\psi(\BV z(t+\varepsilon))]$ of a suitable function
$\psi:\RR^d\to\RR$, for some $\varepsilon>0$. Assume that the
following limit exists:
\begin{equation}
\label{eq:t_limit}
\lim_{\varepsilon\to 0}\frac{\EE[\psi(\BV z(t+\varepsilon)]-\psi(\BV
  z)}\varepsilon=\left.\parderiv{}\varepsilon\EE_\varepsilon[\psi](\BV
z)\right|_{\varepsilon=0}=(\cL\psi)(\BV z),
\end{equation}
where the (generally, integro-differential) linear operator $\cL$,
which is independent of $\psi$, is called the infinitesimal
generator. Clearly, $\cL$ describes the underlying dynamics of the
process -- if $\BV z(t)$ is the phase state of the system of
particles/spheres containing their coordinates and velocities, then
$\cL$ describes the interactions between the spheres. For example, for
the point particles interacting via a long-range potential \cite{Vla},
$\cL$ includes the gradient of the potential function of the force
field.

With $\cL$ specified explicitly, it is easy to see that the forward
equation for the probability density $F(t,\BV z)$ can be derived in a
straightforward fashion \cite{GikSko,App} with the help of the adjoint
operator $\cL^\dagger$, assuming that the latter can also be
explicitly obtained. Integrating the above identity over $\RR^d$
against $F(t,\BV z)\dif\BV z$, we have
\begin{equation}
\int F(t,\BV z)\left(\left.\parderiv{}\varepsilon\EE_\varepsilon[
  \psi](\BV z)\right|_{\varepsilon=0}-(\cL\psi)(\BV z)\right)\dif\BV
z=0.
\end{equation}
First, it can be shown \cite{GikSko} that
\begin{equation}
\label{eq:psi_F}
\int F(t,\BV z)\left.\parderiv{}\varepsilon\EE_\varepsilon[\psi](\BV
z)\right|_{\varepsilon=0}\dif\BV z=\int\psi(\BV z)\parderiv{}t F(t,\BV
z)\dif\BV z.
\end{equation}
Next, with help of the adjoint operator $\cL^\dagger$, we write
\begin{equation}
\int F(t,\BV z)(\cL\psi)(\BV z)\dif\BV z=\int\psi(\BV z)(\cL^\dagger
F)(t,\BV z)\dif\BV z.
\end{equation}
Combining the last two identities and stripping the $\psi$-integral,
we arrive at the general form of the forward Kolmogorov equation
\cite{GikSko,Oks} (also known as the Fokker--Planck equation
\cite{Ris}) for $F$ alone:
\begin{equation}
\label{eq:Lforward}
\parderiv{}t F(t,\BV z)-(\cL^\dagger F)(t,\BV z)=0.
\end{equation}
Since the infinitesimal generator $\cL$ is independent of $F$, its
adjoint $\cL^\dagger$ is also independent of $F$. Thus, the form of
the collision integral in the corresponding BBGKY hierarchy will be
defined entirely by $\cL^\dagger$. This, in turn, will allow to
replace $F$ with a suitable closure without violating any prior
assumptions on the integrand of the collision integral.

Below we introduce the infinitesimal generator into the hard sphere
collision dynamics by randomizing the times at which the velocity
jumps occur. More specifically, the velocity jumps will still occur
according to \eqref{eq:vwxy}, except that, rather then lying precisely
at the collision surface, the coordinate $(\BV x,\BV y)$ of the jump
will be selected at random along the trajectory from within a small
``tolerance'' interval around the collision surface. In such a case,
the conditional expectation $\EE_t[\psi]$ will be a differentiable
function of $t$, even though the velocities of the spheres will still
change instantaneously.  Since a possibility arises that the spheres
become overlapped, we will make provisions for their unimpeded
subsequent separation. The aforementioned tolerance interval can be
made as small as necessary, and the probability of the jump can be
made as large as necessary, so as to mimic the deterministic
collisions with prescribed accuracy. Also, Loschmidt's objection
\cite{Los} will be avoided since the underlying dynamics will be
inherently irreversible.

One can argue that the process we are to propose is a pure abstraction
-- clearly, the actual atoms in a gas are not known to interact
randomly.  However, note that the ``hard sphere collision'' is also an
abstraction -- in reality, no observable physical object can change
its velocity instantaneously. In particular, the atoms in a real gas
do not collide with each other instantaneously, but instead interact
via the Lennard-Jones potential \cite{Len}. Therefore, neither the
deterministic hard spheres, nor our random process are completely
accurate models of molecular interaction on a microscopic level. What
defines the quality of a model here is its ability to describe
macroscopic effects in a gas while being mathematically and logically
consistent with its own abstract foundation.

\section{Random dynamics of hard spheres}
\label{sec:random}

Here we present the details of the random dynamical system which
models collisions of hard spheres, and at the same time possesses the
infinitesimal generator. As before, first we consider the dynamics of
two spheres with coordinates $\BV x$ and $\BV y$, and velocities $\BV
v$ and $\BV w$, respectively. The dynamics of coordinates are, of
course, given via \eqref{eq:xyt}. For the dynamics of velocities, we
consider two separate configurations:
\begin{enumerate}
\item {\bf Collision configuration.} The collision configuration is
  given by the following two criteria, which must hold concurrently:
  \begin{enumerate}
  \item The distance $\|\BV x-\BV y\|$ between the centers of spheres
    satisfies
    \begin{equation}
      \label{eq:contact_zone}
      1-\alpha<\frac{\|\BV x-\BV y\|}\sigma<1+\alpha,
    \end{equation}
    where $0<\alpha\ll 1$ is a constant parameter. The condition above
    signifies that $\|\BV x-\BV y\|/\sigma\approx 1$, within
    $\pm\alpha$-tolerance (that is, the spheres are separated
    approximately by their diameter). We will say that two spheres are
    in the ``contact zone'' whenever the condition in
    \eqref{eq:contact_zone} holds for the coordinates of their
    centers.
    \item The distance $\|\BV x-\BV y\|$ between the centers of
      spheres also diminishes in time, that is,
      \begin{equation}
      \label{eq:incident_spheres}
        (\BV x-\BV y)\cdot(\BV v-\BV w)<0.
      \end{equation}
      This condition signifies that the spheres are approaching each
      other.
  \end{enumerate}
  In the collision configuration, the velocities $(\BV v,\BV w)$ may
  be randomly transformed according to \eqref{eq:vwxy}, with a
  specified probability.  For now, we write informally that in the
  collision configuration the velocities evolve according to
  \begin{equation}
    \dif\BV v(t)=-\dif\BV w(t)=\text{random jump process},
  \end{equation}
  which changes the velocities instantaneously according to
  \eqref{eq:vwxy}.  The jump process must be random for the
  expectation of a jump to be a continuous function of $t$, despite
  the fact that the velocity jumps themselves remain instantaneous.
  The continuity of the expectation is the key property which allows
  the process to possess the infinitesimal generator. The exact
  representation of the requisite jump process will be provided below.
\item {\bf Free-flight configuration.} If the two conditions above in
  \eqref{eq:contact_zone} and \eqref{eq:incident_spheres} do not hold,
  then the spheres are in the free-flight configuration. In this case,
  the velocities are constant in time:
  \begin{equation}
    \dif\BV v(t)=-\dif\BV w(t)=0.
  \end{equation}
\end{enumerate}
At this point, we need to choose a appropriate type of the random jump
process, which will be used as a ``trigger'' to change the velocities
instantaneously. The simplest process which is suitable for the given
task is the {\em point process} \cite{DalVer} -- that is, a scalar,
piecewise-constant random process which starts at 0 and occasionally
increments itself by 1, randomly and independently of previous
increments. Before describing the random hard sphere process in
detail, we formulate a general dynamical process driven by a point
process, and compute the explicit form of its infinitesimal generator.

\subsection{A dynamical system driven by an inhomogeneous point
process}

Here we follow the theory in Section 7.4 of \cite{DalVer} (the
original results are presented in \cite{Papa}). Let $(\Omega,\cF,\PP)$
be the probability space, equipped with a filtration $\{\cF_t\}$,
$t\in\RR_{\geq 0}$, such that for $0\leq t_1\leq t_2<\infty$, the
corresponding sigma-algebras are nested as
$\cF_{t_1}\subseteq\cF_{t_2}\subseteq\cF$. Let $h:\RR_{\geq
  0}\to\RR_{>0}$ be a bounded, strictly positive random variable
adapted to $\{\cF_t\}$. Let $m:\RR_{\geq 0}\to\ZZ_{\geq 0}$ be a
$\{\cF_t\}$-adapted inhomogeneous point process with the conditional
intensity $h(t)$, and the compensator $\tau(t)$, given via
\begin{equation}
\label{eq:taut}
\tau(t)=\int_0^t h(s)\dif s.
\end{equation}
By choosing the random variable $h(t)$ appropriately, we can regulate
the ``temporal density'' of jump points of $m(t)$; indeed, depending
on the magnitude of $h(t)$, the jump points of $m(t)$ can either
arrive in a statistically rapid succession, or, to the contrary,
disperse farther away from each other. For what is to follow, it is a
necessary requirement that the intensity $h(t)$ is a random variable
-- both $m(t)$ and $h(t)$ are, however, adapted to the same filtration
$\{\cF_t\}$, so that if the sequence of values of $h$ up to $t$ is
given, so is the sequence of values $m$.

We now denote the corresponding jump process of $m(t)$ via
$\Delta m(t)$:
\begin{equation}
\Delta m(t)=m(t)-m(t-).
\end{equation}
Above, the notation ``$t-$'' denotes the left-limit at $t$.  Let
$M(t,\cdot)$ be the corresponding random measure of $\Delta
m(t)$:
\begin{equation}
M(t,A)=\text{number of values of }\Delta m(s)\in A\subset
\RR_{>0}\text{ in }0<s\leq t.
\end{equation}
As $\Delta m(t)$ only assumes the values of either 0 or 1,
$M(t,\cdot)$ is concentrated at 1.

Let us now define a stochastic process $\BV z(t)$ on a
Euclidean space $\RR^d$ as follows:
\begin{equation}
\label{eq:stoch_proc}
\BV z(t)=\BV z(0)+\int_0^t\BV f(s)\dif s+\int_0^t\int_{\RR_{>0}}\xi
\BV g(s) M(\dif s,\dif\xi),
\end{equation}
where $\BV f,\BV g:\RR_{\geq 0}\to\RR^d$ are suitable (for the purpose
of stochastic integration above) random variables, adapted to
$\{\cF_t\}$. Clearly, $\BV z(t)$ is also $\{\cF_t\}$-adapted by
construction.

Our task here is to compute the infinitesimal generator of $\BV z(t)$,
that is, for a test function $\psi(\BV z)$, we would like to compute
\begin{equation}
\cL[\psi](t)=\lim_{\varepsilon\to 0}\frac 1\varepsilon \big(\EE[\psi(\BV
  z(t+\varepsilon))|\cF_t]-\psi(\BV z(t))\big).
\end{equation}
To compute the expectation above, we need to adapt the integral form
of $\BV z$ in \eqref{eq:stoch_proc} to the It\^o formula for
L\'evy-type stochastic integrals (see Chapter 4 of \cite{App}).
However, the problem is that the random measure in the right-hand side
of \eqref{eq:stoch_proc} is not that of the standard Poisson point
process with constant intensity, but that of the point process with
random intensity $h(t)$, defined above.

Our first step here is, therefore, the transformation of the
stochastic integral in the right-hand side of \eqref{eq:stoch_proc} to
an integral against the random measure of a standard Poisson point
process. According to Theorem 7.4.I of \cite{DalVer} (also see
\cite{Papa}), the random point process $n:\RR_{\geq 0}\to\ZZ_{\geq
  0}$, defined via
\begin{equation}
n(t)=m(\tau^{-1}(t)),
\end{equation}
is the standard Poisson point process with intensity 1, where
$\tau(t)$ is the random, albeit $\{\cF_t\}$-adapted, compensator
process of $m(t)$, defined above in \eqref{eq:taut}. Subsequently, we
can write the stochastic integral in \eqref{eq:stoch_proc} from $t$ to
$t+\varepsilon$ as
\begin{multline}
  \int_t^{t+\varepsilon}\int_{\RR_{>0}}\xi\BV g(s) M(\dif s,\dif\xi)=
  \sum_{\myatop{t<s\leq t+\varepsilon}{\Delta m(s)>0}}\Delta m(s)\BV
  g(s)=\sum_{\myatop{t<s\leq t+\varepsilon}{\Delta n(\tau(s))>0}}
  \Delta n(\tau(s))\BV g(s) =\\= \sum_{\myatop{\tau(t)<s\leq\tau(t+
      \varepsilon) }{\Delta n(s)>0 }}\Delta n(s)\BV g(\tau^{-1}(s))=
  \int_{\tau(t) }^{\tau(t+\varepsilon)} \int_{\RR_{>0}}\xi\BV
  g(\tau^{-1}(s)) N(\dif s,\dif\xi),
\end{multline}
with $N(t,\cdot)$ being the random measure of the standard Poisson
point process with intensity 1; in particular, its intensity measure
$\EE N$ \cite{App} is given via
\begin{equation}
\EE N(\dif s,\dif\xi)=\delta(\xi-1)\dif\xi\dif s.
\end{equation}
Substituting the above integral into \eqref{eq:stoch_proc}, we write
\begin{equation}
\BV z(t+\varepsilon)=\BV z(t)+\int_t^{t+\varepsilon}\BV f(s)\dif s+
\int_{\tau(t)}^{\tau(t+\varepsilon)} \int_{\RR_{>0}}\xi\BV
g(\tau^{-1}(s)) N(\dif s,\dif\xi).
\end{equation}
Next, recalling the It\^o formula for L\'evy-type stochastic integrals
in Chapter 4 of \cite{App}, we write
\begin{multline}
\psi(\BV z(t+\varepsilon))-\psi(\BV z(t))= \int_t^{t+\varepsilon}
\psi'(\BV z(s-))\BV f(s)\dif s+\\+\int_{\tau(t)}^{\tau(t+\varepsilon)}
\int_{\RR_{>0}}\left(\psi\big(\BV z(\tau^{-1}(s)-)+\xi\BV g(\tau^{-1}
(s))\big)-\psi(\BV z(\tau^{-1}(s)-))\right)N(\dif s,\dif\xi).
\end{multline}
Applying the conditional expectation on both sides, we obtain, for the
left-hand side and the first term in the right-hand side,
\begin{subequations}
\begin{equation}
\EE[\psi(\BV z(t+\varepsilon))-\psi(\BV z(t))|\cF_t]=\EE[\psi(\BV
  z(t+\varepsilon))|\cF_t]-\psi(\BV z(t)),
\end{equation}
\begin{equation}
\EE\int_t^{t+\varepsilon} \psi'(\BV z(s-))\BV f(s)\dif s= \EE\int_0
^\varepsilon \psi'(\BV z(t+s-))\BV f(t+s)\dif s=\varepsilon\psi'(\BV
z(t))\BV f(t)+o(\varepsilon),
\end{equation}
\end{subequations}
where in the second expression the conditional expectation disappears
in the leading order term because the $\BV z(t)$ and $\BV f(t)$ are
both $\{\cF_t\}$-adapted.

The expectation of the stochastic integral is somewhat more
complicated.  First, observe that
\begin{equation}
\tau(t+\varepsilon)=\tau(t)+\varepsilon h(t)+o(\varepsilon),
\end{equation}
where in the right-hand side both leading order terms are
$\{\cF_t\}$-adapted. The integral then can be expressed as
\begin{multline}
\int_{\tau(t)}^{\tau(t+\varepsilon)} \int_{\RR_{>0}}\left(\psi\big(\BV
z(\tau^{-1}(s)-)+\xi\BV g(\tau^{-1} (s))\big)-\psi(\BV z(\tau^{-1}
(s)-))\right)N(\dif s,\dif\xi)=\\=\int_{\tau(t)}^{\tau(t)+\varepsilon
  h(t)} \int_{\RR_{>0}}\left(\psi\big(\BV z(\tau^{-1}(s)-)+\xi\BV
g(\tau^{-1} (s))\big)-\psi(\BV z(\tau^{-1} (s)-))\right)N(\dif
s,\dif\xi)+o(\varepsilon)=\\=\int_{\tau(t)}^{\tau(t)+\varepsilon h(t)}
\int_{\RR_{>0}}\left(\psi\big(\BV z(t-)+\xi\BV g(t)\big)-\psi(\BV
z(t-))\right)N(\dif s,\dif\xi)+o(\varepsilon),
\end{multline}
where in the second identity the dummy variable of integration $s$ was
replaced with its starting value $\tau(t)$.

At this point, observe that in the leading order integral above, the
limits of integration, as well as the integrand, are
$\{\cF_t\}$-adapted.  Thus, applying the conditional expectation to
the leading order integral yields
\begin{multline}
\EE\int_{\tau(t)}^{\tau(t)+\varepsilon h(t)} \int_{\RR_{>0}} \left(
\psi\big(\BV z(t-)+\xi\BV g(t)\big)-\psi(\BV z(t-))\right)N(\dif
s,\dif\xi)=\\=\int_{\tau(t)}^{\tau(t)+\varepsilon h(t)}
\int_{\RR_{>0}}\left(\psi\big(\BV z(t)+\xi\BV g(t)\big)-\psi(\BV
z(t))\right)\EE N(\dif s,\dif\xi)=\\=\varepsilon h(t)
\left(\psi\big(\BV z(t)+\BV g(t)\big)-\psi(\BV z(t))\right).
\end{multline}
Assembling the terms together, we obtain the infinitesimal generator
in the form
\begin{equation}
\cL[\psi]=\lim_{\varepsilon\to 0}\frac{\EE[\psi(\BV z(t+\varepsilon
    ))|\cF_t]-\psi(\BV z(t))} \varepsilon=\psi'(\BV z)\BV f(t)+h(t)
\big(\psi(\BV z+\BV g(t))-\psi(\BV z)\big),
\end{equation}
where in the last identity we denote $\BV z=\BV z(t)$, for brevity.

It is worth noting that the form of the infinitesimal generator above
extends naturally onto the free-flight configuration of the dynamics
-- it suffices to set the variable intensity $h(t)=0$ above. In this
case, $\BV z(t)$ in \eqref{eq:stoch_proc} will be driven solely by the
integral over the vector field $\BV f(t)$ alone.

\subsection{Random dynamics of two spheres}

To adapt the general stochastic process in \eqref{eq:stoch_proc} to
the dynamics of spheres, we need to relate $\BV z(t)$, $\BV f(t)$,
$\BV g(t)$ and $h(t)$ to the variables of the dynamics. Obviously,
$\BV z(t)$ is the state vector of the system, and thus it is going to
incorporate the coordinates and velocities of both spheres:
\begin{equation}
\label{eq:z}
\BV z(t)=\left(\begin{array}{l}
\BV x(t) \\ \BV y(t) \\ \BV v(t) \\ \BV  w(t)
\end{array}\right).
\end{equation}
Subsequently, $\BV f(t)$ is related to the deterministic component of
the dynamics, which is the evolution of the coordinates $\BV x(t)$ and
$\BV y(t)$ for given velocities $\BV v$ and $\BV w$ according to
\eqref{eq:xyt}:
\begin{equation}
\label{eq:fz}
\BV f(t)=\BV f(\BV z(t-))=\left(\begin{array}{l}
\BV v(t-) \\ \BV w(t-) \\ \BV 0 \\ \BV 0
\end{array}\right).
\end{equation}
To specify $\BV g(t)$, we observe that the instantaneous change of
velocities in \eqref{eq:vwxy} can be written, with help of the jump
process $\Delta m(t)$, as
\begin{equation}
\label{eq:velocity_jumps}
\BV v(t)-\BV v(t-)=-(\BV w(t)-\BV w(t-))=\Delta m(t)\big(\BV w(t-)-\BV
v(t-) \big)\cdot(\BV x-\BV y)\frac{\BV x-\BV y}{\|\BV x-\BV y\|^2},
\end{equation}
where $\BV x=\BV x(t)$, $\BV y=\BV y(t)$. Therefore, we can define
$\BV g(t)$ via
\begin{equation}
\label{eq:gz}
\BV g(t)=\BV g(\BV z(t-))=\frac{(\BV x-\BV y)\cdot(\BV v(t-)-\BV
  w(t-)) }{\|\BV x-\BV y\|^2}\left(\begin{array}{l} \BV 0 \\ \BV 0
    \\ \BV y-\BV x \\ \BV x-\BV y \end{array}\right).
\end{equation}
Then we write the process in \eqref{eq:xyt}+\eqref{eq:velocity_jumps}
as the following stochastic differential equation \cite{App}:
\begin{equation}
\label{eq:dyn_sys_2}
\BV z(t)=\BV z(0)+\int_0^t\BV f(\BV z(s-))\dif s+\int_0^t
\int_{\RR_{>0}}\xi\BV g(\BV z(s-))M(\dif s,\dif\xi),
\end{equation}
with $\BV z$, $\BV f(\BV z)$ and $\BV g(\BV z)$ given via
\eqref{eq:z}, \eqref{eq:fz} and \eqref{eq:gz}, respectively.

It remains to specify the variable intensity $h(t)$, which should
activate the point process $m(t)$ when both \eqref{eq:contact_zone}
and \eqref{eq:incident_spheres} hold concurrently, and be zero in the
free-flight configuration.  Here, we define $h(t)$ as
\begin{equation}
\label{eq:ht}
h(t)=h(\BV z(t-))=\lambda\Hside\big((\BV x-\BV y)\cdot(\BV w-\BV
v)\big)\delta_{\alpha\sigma}(\|\BV x-\BV y\|-\sigma)\frac{\BV x-\BV
  y}{\|\BV x-\BV y\|}\cdot(\BV w-\BV v).
\end{equation}
Above, $\lambda>0$, $0<\alpha\ll 1$ are constant parameters, and
$\delta_\alpha(x)$ is the standard mollifier of the delta-function
$\delta(x)$, given via
\begin{equation}
\label{eq:mollifier}
\delta_\alpha(x)=\frac 1\alpha\phi\left(\frac x\alpha\right),\qquad
\phi(x)=\left\{\begin{array}{l@{\quad}l} ce^{-\frac 1{1-x^2}}, & |x|<
1, \\ 0, & |x|\geq 1,\\\end{array}\right.\qquad\int_{-1}^1\phi(x)\dif
x=1,
\end{equation}
where the constant parameter $c$ ensures the proper normalization. For
a function $\psi(x)$, which is continuous at zero, we thus have
\begin{equation}
\lim_{\alpha\to 0}\int_{-\infty}^{+\infty}\psi(x)\delta_\alpha(x)\dif
x=\psi(0),
\end{equation}
that is, $\delta_\alpha$ can serve as the delta-function in the limit
$\alpha\to 0$, while remaining smooth for finitely small $\alpha$. We
will denote the anti-derivative of $\delta_\alpha$ as $\Hside_\alpha$:
\begin{equation}
\Hside_\alpha(x)=\int_{-\infty}^x\delta_\alpha(y)\dif y.
\end{equation}
Clearly, as $\alpha\to 0$, $\Hside_\alpha(x)$ becomes the usual
Heaviside step-function.

Observe that, in the collision configuration
\eqref{eq:contact_zone}--\eqref{eq:incident_spheres}, the variable
intensity of the point process in \eqref{eq:ht}, as required in
\cite{DalVer,Papa}, is indeed $\cF_t$-adapted, strictly positive and
bounded, since, first, $\BV z(t)$ is $\cF_t$-adapted by construction,
and, second, whenever \eqref{eq:contact_zone} and
\eqref{eq:incident_spheres} hold, we have
\begin{equation}
0< h(\BV z)\leq\lambda\delta_{\alpha\sigma}(0)\|\BV w-\BV v\|\leq 2
\lambda\delta_{\alpha\sigma}(0)\sqrt E,
\end{equation}
where $E$ is the constant energy of the system of two spheres.

The dynamics in \eqref{eq:dyn_sys_2}--\eqref{eq:ht} function as
follows:
\begin{itemize}
\item If $\|\BV x-\BV y\|$ is away from $\sigma$, or if $\|\BV x-\BV
  y\|$ is growing in time (that is, the spheres are escaping each
  other), then the triggering point process is not present, and the
  spheres are moving with constant velocities according to
  \eqref{eq:dyn_sys}. Accordingly, the intensity $h(\BV z(t))$ in
  \eqref{eq:ht} is zero in the infinitesimal generator of the process,
  and thus the generator consists of the free-flight term only.
\item Once $\|\BV x-\BV y\|$ is close enough to $\sigma$ and, at the
  same time $\|\BV x-\BV y\|$ is decreasing in time, the spheres enter
  the contact zone (both conditions \eqref{eq:contact_zone} and
  \eqref{eq:incident_spheres} are satisfied). In this case, the
  collision-triggering point process becomes present, with the
  intensity $h(\BV z(t))$ being strictly greater than zero. Then,
  there are two possibilities:
\begin{itemize}
\item A jump in the point process arrives so that the spheres
  ``collide'' according to \eqref{eq:vwxy}. In this case, the spheres
  start escaping each other (so that \eqref{eq:incident_spheres} no
  longer holds) and the triggering point process is no longer
  present. In the infinitesimal generator, the Heaviside function
  becomes zero and so does the intensity $h(\BV z(t))$.
\item A jump does not arrive, so that eventually
  $\delta_{\alpha\sigma}(\|\BV x-\BV y\|-\sigma)$ decays back to zero
  together with the intensity $h(\BV z(t))$ of the point process; in
  this case, the spheres pass through each other without interaction.
\end{itemize}
In either scenario, the point process $m(t)$ becomes dormant until the
spheres approach each other again.
\end{itemize}
The existence of strong solutions to
\eqref{eq:dyn_sys_2}--\eqref{eq:ht} in the collision configuration
\eqref{eq:contact_zone}--\eqref{eq:incident_spheres} is a subject that
merits a separate discussion. For the purpose of this work, we will
assume that bounded strong solutions are sufficiently generic for
typical initial conditions, so that the corresponding statistical
formulation (the forward Kolmogorov equation) of the dynamics is
reliable enough for the description of large ensembles of solutions.

The definition of the jump intensity above in \eqref{eq:ht} also
indicates that the probability that the jump in the point process does
not arrive during the collision window is $e^{-\lambda}$, regardless
of the values of $\sigma$ or $\alpha$. Indeed, observe that one can
write
\begin{equation}
\label{eq:ht2}
h(t)=\lambda\Hside\left(-\deriv{}t\|\BV x(t)-\BV y(t)\|\right)
\deriv{}t\Hside_{\alpha\sigma}\big(\sigma-\|\BV x(t)-\BV y(t)\| \big),
\end{equation}
which means that if the contact zone is traversed completely (that is,
the jump has not arrived), then the compensator $\tau(t)$ in
\eqref{eq:taut} is incremented by $\lambda$.  Clearly, to mimic the
collisions of hard deterministic spheres in
Section~\ref{sec:conventional}, one eventually needs to take
$\alpha\to 0$ and $\lambda\to\infty$, so that, first, the contact zone
\eqref{eq:contact_zone} becomes infinitely thin, and, second, the jump
arrives with probability 1 whenever the spheres are in the collision
configuration.

The corresponding infinitesimal generator of
\eqref{eq:dyn_sys_2}--\eqref{eq:ht} is given via
\begin{equation}
\cL[\psi]=\BV f(\BV z)\cdot\parderiv{\psi}{\BV z}+h(\BV z)
\big(\psi(\BV z+\BV g(\BV z))-\psi(\BV z)\big).
\end{equation}
Changing back to the original variables $\BV x$, $\BV y$,
$\BV v$ and $\BV w$, we write the infinitesimal generator of
\eqref{eq:dyn_sys_2} in the form
\begin{multline}
\label{eq:generator}
\cL[\psi]=\BV v\cdot\parderiv\psi{\BV x}+\BV w\cdot \parderiv\psi{\BV
  y}+\lambda\delta_{\alpha\sigma}(\| \BV x-\BV y\|- \sigma)\frac{\BV
  x-\BV y}{\|\BV x-\BV y\|}\cdot(\BV w-\BV v)\\\Hside\big((\BV x-\BV
y)\cdot(\BV w-\BV v)\big)\big(\psi(\BV x,\BV y,\BV v',\BV w')-\psi(\BV
x,\BV y,\BV v,\BV w)\big),
\end{multline}
where $\BV v'$ and $\BV w'$ are the functions of $\BV x$, $\BV y$,
$\BV v$ and $\BV w$ given in~\eqref{eq:vwxy}. The jump portion of the
generator above in \eqref{eq:generator} is not translationally
invariant, and thus the process in \eqref{eq:dyn_sys_2} is not a
L\'evy process. However, it is a L\'evy-type Feller process
\cite{App,Fel2}, whose infinitesimal generator can be reformulated in
the Courr\`ege form \cite{Cou} via an appropriate change of variables.

The next step is to obtain the corresponding forward Kolmogorov
equation \cite{GikSko,Oks,Ris} for the probability density of states
of the system, which is easily achieved via the integration by
parts. Let $F(t,\BV x,\BV y,\BV v,\BV w)$ be the corresponding
probability distribution of the random process above. We can then
integrate~\eqref{eq:generator} against $F$ and obtain, with help of
\eqref{eq:psi_F},
\begin{multline}
\int\bigg(\psi\parderiv Ft-F\BV v\cdot\parderiv\psi{\BV x}-F\BV w\cdot
\parderiv\psi{\BV y}-F\lambda\big(\psi(\BV x,\BV y,\BV v',\BV w')-
\psi(\BV x,\BV y,\BV v,\BV w)\big)\\\frac{\BV x-\BV y}{\| \BV x-\BV
  y\|}\cdot(\BV w-\BV v)\Hside\big((\BV x-\BV y)\cdot(\BV w- \BV
v)\big)\delta_{\alpha\sigma}(\|\BV x-\BV y\|-\sigma)\bigg)\dif V_2
\dif S_2=0,
\end{multline}
where $\dif V_2$ is the volume element of the coordinate space, and
$\dif S_2$ is the area element of the sphere of zero momentum and
constant energy (the subscript denotes the number of spheres in the
system).

Above, the terms with spatial derivatives in $\BV x$ and $\BV y$ can
be integrated by parts, with the condition that the boundary effects
are not present. For the part with $\psi(\BV v',\BV w')$ we can write,
for fixed $\BV x$ and $\BV y$,
\begin{multline}
\int\psi(\BV x,\BV y,\BV v',\BV w')F(\BV x,\BV y,\BV v,\BV w)\frac{\BV
  x-\BV y}{\|\BV x-\BV y \|}\cdot(\BV w-\BV v)\Hside\big((\BV x-\BV y)
\cdot(\BV w-\BV v)\big)\dif S_2=\\=\int\psi(\BV x,\BV y,\BV v',\BV w')
F(\BV x,\BV y,\BV v,\BV w)\frac{\BV x-\BV y}{\|\BV x-\BV y\|}\cdot(
\BV v'-\BV w')\Hside\big((\BV x-\BV y)\cdot (\BV v'-\BV w')\big)\dif
S_2=\\=-\int\psi(\BV x,\BV y,\BV v,\BV w)F(\BV x,\BV y,\BV v',\BV w')
\frac{\BV x-\BV y}{\|\BV x -\BV y\|}\cdot(\BV w-\BV v)\Hside\big((\BV
x-\BV y)\cdot(\BV v-\BV w)\big)\dif S_2,
\end{multline}
where we used \eqref{eq:vwxyn}, \eqref{eq:vwxy_inv}
and~\eqref{eq:det1} (note that $\BV v'$ and $\BV w'$ remain on the
same zero momentum -- constant energy sphere), and in the last
identity renamed $\BV v'\to\BV v$, $\BV w'\to\BV w$ and vice versa,
since the integral occurs over the same velocity sphere. As a result,
we can recombine the terms as
\begin{multline}
\int\psi\bigg(\parderiv Ft+\BV v\cdot\parderiv F{\BV x}+\BV w\cdot
\parderiv F{\BV y}+\lambda\delta_{\alpha\sigma}(\|\BV x-\BV y\|-\sigma
)\frac{\BV x-\BV y}{\|\BV x-\BV y\|}\cdot(\BV w-\BV v)\\\left[F(\BV
  x,\BV y,\BV v,\BV w) \Hside \big((\BV x-\BV y)\cdot(\BV w-\BV v)
  \big)+F(\BV x, \BV y,\BV v',\BV w')\Hside\big((\BV x-\BV y)\cdot
  (\BV v-\BV w)\big) \right]\bigg)\dif V_2\dif S_2=0.
\end{multline}
Assuming that $\psi$ is arbitrary, we can strip the integral over
$\psi$ and obtain the equation for $F$ alone:
\begin{multline}
\label{eq:kolmogorov2}
\parderiv Ft+\BV v\cdot\parderiv F{\BV x}+\BV w\cdot\parderiv F{\BV y}
+\lambda\delta_{\alpha\sigma}(\|\BV x-\BV y\|-\sigma)\frac{\BV x-\BV y
}{\|\BV x-\BV y\|}\cdot(\BV w-\BV v)\\\left[F(\BV x,\BV y,\BV v,\BV w)
  \Hside\big((\BV x-\BV y)\cdot(\BV w-\BV v)\big)+F(\BV x,\BV y,\BV
  v',\BV w')\Hside \big((\BV x-\BV y)\cdot(\BV v-\BV w)\big)\right]
=0.
\end{multline}
Unlike the Liouville problem \eqref{eq:F_zero}--\eqref{eq:BC}, here
observe that the effect of collisions is present in the equation
itself, and is not contingent upon additional properties imposed on
$F$.

\subsection{Extension to many spheres}

Here we extend the previously formulated dynamics onto $K$ spheres,
with the corresponding coordinates $\BV x_i(t)$ and velocities $\BV
v_i(t)$, $1\leq i\leq K$. Observe that we have $K(K-1)/2$ possible
pairs of spheres. In order to define their random interactions, we
introduce $K(K-1)/2$ independent instances of the point process, each
assigned to the pair of $i$-th and $j$-th spheres.

For $1\leq i<j\leq K$, let us define
\begin{equation}
\BV Z=(\BV x_1,\ldots,\BV x_K,\BV v_1,\ldots,\BV v_K)^T,\quad\BV
z_{ij}=(\BV x_i,\BV x_j,\BV v_i,\BV v_j)^T,\quad
\BV F(\BV Z)=(\BV v_1,\ldots,\BV v_K,\BV 0,\ldots,\BV 0)^T,
\end{equation}
\begin{equation}
\BV G_{ij}(\BV Z)=\frac{(\BV x_i-\BV x_j)\cdot(\BV v_i-\BV v_j)}{\|\BV
  x_i- \BV x_j\|^2}(\BV 0,\ldots,\BV 0,\BV x_j-\BV x_i,\BV
0,\ldots,\BV 0,\BV x_i-\BV x_j,\BV 0,\ldots,\BV 0)^T,
\end{equation}
where the two nonzero entries in the $\BV G_{ij}$-vector above are in
the $(K+i)$-th and $(K+j)$-th slots. Let $m_{ij}(t)$ be the set of
$K(K-1)/2$ independent inhomogeneous point processes with conditional
intensities $h_{ij}(t)$, given via
\begin{equation}
  h_{ij}(t)=h(\BV z_{ij}(t)),
\end{equation}
where $h(\BV z)$ is defined in \eqref{eq:ht}. Let $M_{ij}(t,\cdot)$ be
the set of corresponding random measures for $m_{ij}(t)$. Then, the
$K$-sphere dynamics is defined via the following system of stochastic
differential equations:
\begin{equation}
\label{eq:dyn_sys}
\BV Z(t)=\BV Z(0)+\int_0^t\BV F(\BV Z(s-))\dif s+\sum_{i=1}^{K-1}
\sum_{j=k+1}^K\int_0^t\int_{\RR_{>0}}\xi\BV G_{ij}(\BV Z(s-)) M_{ij}
(\dif s,\dif\xi).
\end{equation}
The process above in~\eqref{eq:dyn_sys} is also a L\'evy-type Feller
process \cite{App,Fel2}, which lives on the sphere of zero momentum
and constant energy
\begin{equation}
\label{eq:mom_en}
\sum_{i=1}^K\BV v_i=\BV 0,\qquad E=\frac 12\sum_{i=1}^K\|\BV v_i\|^2.
\end{equation}
As above for two spheres, here we assume that the total momentum of
the system is zero without loss of generality.  The infinitesimal
generator of such process is, apparently, given via
\begin{equation}
\cL[\psi]=\BV F(\BV Z)\cdot\parderiv{\psi}{\BV Z}+\sum_{i=1}^{K-1}
\sum_{j=i+1}^K h(\BV z_{ij})\big(\psi(\BV Z+\BV G_{ij}(\BV Z))
-\psi(\BV Z)\big).
\end{equation}
In the $\BV x_i$ and $\BV v_i$ variables, this translates into
\begin{multline}
\label{eq:generator_K}
\cL[\psi]=\sum_{i=1}^K\BV v_i\cdot \parderiv\psi{\BV x_i}+\sum_{i=1}
^{K-1}\sum_{j=i+1}^K\lambda\delta_{ \alpha \sigma}(\|\BV x_i-\BV x_j\|
-\sigma)\frac{\BV x_i-\BV x_j}{\| \BV x_i-\BV x_j\|}\cdot(\BV v_j-\BV
v_i)\\\Hside\big((\BV x_i-\BV x_j)\cdot(\BV v_j-\BV v_i)\big)\big(\psi
(\BV v_i',\BV v_j') -\psi\big),
\end{multline}
where the notation $\psi(\BV v_i',\BV v_j')$ specifies that all
velocity arguments in $\psi$ are set to the corresponding velocities
$\BV v_k$, except for $i$-th and $j$-th, which are set to $\BV v_i'$
and $\BV v_j'$ via \eqref{eq:vwxy}.

Observe that the dynamics in \eqref{eq:dyn_sys} is a direct extension
of the dynamics of two spheres in \eqref{eq:dyn_sys_2} onto multiple
spheres -- the evolution of the coordinates is governed by the same
equations, and the velocities of each sphere are coupled to all other
spheres via the independent point processes $m_{ij}(t)$. In
particular, there is no provision for a collision of more than two
spheres at once (which is often discussed in the literature
\cite{Cer,CerIllPul}); however, given the fact that the collisions in
the hard sphere model are instantaneous, we will assume that the event
of a three-sphere collision is improbable for a ``generic'' initial
condition.

It is interesting that the properties of an infinitesimal generator
similar to \eqref{eq:generator_K} were studied in \cite{Rez} in the
same context (that is, a system of $K$ particles interacting according
to \eqref{eq:vwxy}). However, the collision part of the infinitesimal
generator in \cite{Rez} was scaled differently, as if the intensity
parameter $\lambda$ in \eqref{eq:generator_K} was set to
$(\alpha\sigma)^3$. Thus, as $\alpha\to 0$, the particles described by
the generator in \cite{Rez} ceased colliding upon contact. It is,
however, unclear where the infinitesimal generator in \cite{Rez} comes
from; the generator in \cite{Rez} appears to be postulated, rather
than derived from an underlying SDE.

To obtain the corresponding forward equation, we follow the same
principle as for the two spheres above.  First, we integrate
\eqref{eq:generator_K} against the probability density $F$ and obtain
\begin{multline}
\int\bigg(\psi\parderiv Ft+\psi\sum_{i=1}^K\BV v_i\cdot\parderiv F{\BV
  x_i}-F\sum_{i=1}^{K-1}\sum_{j=i+1}^K\lambda\big(\psi(\BV v_i',\BV
v_j') -\psi\big)\\\frac{\BV x_i-\BV x_j}{\|\BV x_i-\BV x_j\|}\cdot(\BV
v_j-\BV v_i)\Hside\big((\BV x_i-\BV x_j)\cdot(\BV v_j-\BV v_i)\big)
\delta_{\alpha\sigma}(\|\BV x_i-\BV x_j\|-\sigma)\bigg)\dif V_K\dif
S_K=0,
\end{multline}
where $\dif V_K$ is the volume element of the coordinate space of the
$K$ spheres, $\dif S_K$ is the area element of the corresponding
velocity sphere of zero momentum and constant energy, and the term
with the spatial derivatives is integrated by parts assuming that the
boundary terms vanish.  Now, for all terms with $\BV v_i'$ and $\BV
v_j'$ we have, for fixed coordinates,
\begin{multline}
\int F(\BV v_i,\BV v_j)\psi(\BV v_i',\BV v_j')\frac{\BV x_i-\BV x_j}{
  \|\BV x_i-\BV x_j\|}\cdot(\BV v_j-\BV v_i)\Hside\big((\BV x_i-\BV
x_j)\cdot(\BV v_j-\BV v_i)\big)\dif S_K=\\=\int F(\BV v_i,\BV v_j)
\psi(\BV v_i',\BV v_j')\frac{\BV x_i-\BV x_j}{\|\BV x_i-\BV x_j\|}
\cdot(\BV v_i'-\BV v_j')\Hside\big((\BV x_i-\BV x_j)\cdot(\BV v_i'
-\BV v_j')\big)\dif S_K=\\=-\int F(\BV v_i',\BV v_j')\psi(\BV v_i,\BV
v_j)\frac{\BV x_i-\BV x_j}{\|\BV x_i-\BV x_j\|}\cdot(\BV v_j-\BV
v_i)\Hside\big((\BV x_i-\BV x_j)\cdot(\BV v_i-\BV v_j)\big)\dif S_K,
\end{multline}
where we used \eqref{eq:vwxyn}, \eqref{eq:vwxy_inv}
and~\eqref{eq:det1}, and observed that, for fixed coordinates, the
variables $\BV v_i'$ and $\BV v_j'$ sample the same zero momentum and
constant energy sphere as do $\BV v_i$ and $\BV v_j$. Finally,
stripping the integral over $\psi\dif V_K\dif S_K$, we arrive at the
forward Kolmogorov equation in the form
\begin{multline}
\label{eq:kolmogorov}
\parderiv Ft+\sum_{i=1}^K\BV v_i\cdot\parderiv F{\BV x_i}+\sum_{i=1 }
^{K-1}\sum_{j=i+1}^K\lambda\delta_{\alpha\sigma}(\|\BV x_i-\BV x_j\|
-\sigma)\frac{\BV x_i-\BV x_j}{ \|\BV x_i-\BV x_j\|}\cdot(\BV v_j-\BV
v_i)\\\left[F\Hside\big((\BV x_i-\BV x_j)\cdot(\BV v_j-\BV v_i)\big)
  +F(\BV v_i',\BV v_j')\Hside\big((\BV x_i-\BV x_j)\cdot(\BV v_i-\BV
  v_j)\big)\right]=0.
\end{multline}
Note that the equation above in \eqref{eq:kolmogorov} admits solutions
which are symmetric under the reordering of the spheres. These
solutions are ``physical'', that is, they correspond to real-world
scenarios where it is impossible to statistically tell the spheres
apart.

\section{Some properties of the random sphere dynamics}
\label{sec:properties}

\subsection{A two-sphere solution along a characteristic}

Above we implied that sending $\alpha\to 0$ and $\lambda\to\infty$ in
\eqref{eq:dyn_sys_2}--\eqref{eq:ht} should result in a reasonable
approximation of the conventional hard sphere dynamics described in
\eqref{eq:xyt} and \eqref{eq:vwxy}. Here we examine the behavior of
the solutions of the Kolmogorov equation \eqref{eq:kolmogorov2} and
compare it to the behavior of the solutions of
\eqref{eq:F_zero}--\eqref{eq:BC}.

We solve \eqref{eq:kolmogorov2} using the method of characteristics;
namely, we treat the advection term of \eqref{eq:kolmogorov2} as the
ordinary, scalar spatial derivative in the direction of $(1,\BV v,\BV
w)$, with the directional parameter denoted as $s$. With this, $t$,
$\BV x$ and $\BV y$ are given via the straight line
\begin{equation}
(t(s),\BV x(s),\BV y(s))=(0,\BV x_0,\BV y_0)+ s(1,\BV v,\BV w).
\end{equation}
On this straight line, \eqref{eq:kolmogorov2} becomes
\begin{multline}
\deriv Fs=\lambda\deriv{}s\Hside_{\alpha\sigma}(\|\BV x(s)-\BV y(s)\|
-\sigma)\big[F \Hside\big((\BV x(s)-\BV y(s))\cdot(\BV w-\BV v)\big)
  +\\+ F(s,\BV x(s),\BV y(s),\BV v'(s),\BV w'(s))\Hside\big((\BV x(s)
  -\BV y(s))\cdot(\BV v-\BV w)\big)\big].
\end{multline}
Here, the two terms in the right-hand side are never nonzero
simultaneously, due to the Heaviside step-functions. We assume that
the initial condition $(\BV x_0,\BV y_0)$ satisfies $\|\BV x_0-\BV
y_0\|>\sigma(1+\alpha)$, and $(\BV x_0-\BV y_0)\cdot(\BV v-\BV w)<0$,
so that the spheres are not overlapping and on the ``collision
course''. Then, as the characteristic traverses the contact zone for
the first time, we have
\begin{equation}
\deriv Fs=\lambda F\deriv{}s\Hside_{\alpha\sigma}(\|\BV x(s)-\BV y(s)
\|-\sigma),
\end{equation}
which results in the solution
\begin{equation}
F(s)=F_0\exp\left(\lambda\Hside_{\alpha\sigma}(\|\BV x(s)-\BV y(s)\|
-\sigma)-\lambda\Hside_{\alpha\sigma}(\|\BV x_0-\BV y_0\|-\sigma)
\right).
\end{equation}
Observe that $F(s)=F_0$ if $\|\BV x(s)-\BV y(s)\|>\sigma(1+\alpha)$,
and $F(s)=e^{-\lambda}F_0$ if $\|\BV x(s)-\BV y(s)\|<\sigma(1-
\alpha)$.  In the original variables, the solution translates into
\begin{multline}
F(t,\BV x,\BV y,\BV v,\BV w)=F_0(\BV x-t\BV v,\BV y-t\BV w,\BV v,\BV
w)\\\exp\left(\lambda\Hside_{\alpha\sigma}(\|\BV x-\BV y\|-\sigma)-
\lambda\Hside_{\alpha\sigma}(\|\BV x-\BV y-t(\BV v-\BV w)\|-\sigma)
\right).
\end{multline}
Sending $\alpha\to 0$ has no effect other than making the contact zone
infinitely thin, and, subsequently, the transition between $F_0$ and
$e^{-\lambda}F_0$ instantaneous.

As we continue along the characteristic, it eventually traverses the
contact zone again, and exits the overlapped state. In this
configuration, we have
\begin{equation}
\deriv Fs=\lambda F(\BV x(s),\BV y(s),\BV v'(s),\BV w'(s)) \deriv{}s
\Hside_{\alpha\sigma}(\|\BV x(s)-\BV y(s)\|-\sigma).
\end{equation}
Here, the right-hand side contains the values of $F(\BV v',\BV w')$,
which are carried along the characteristic curves with directions
given via
\[\BV v'(s)=\BV v'(\BV x(s),\BV y(s),\BV v,\BV w),\qquad\BV
w'(s)=\BV w'(\BV x(s),\BV y(s),\BV v,\BV w),\]
and should be treated as an external forcing. However, we already know
from what we found above that we can express, for some $t^*$,
\begin{multline}
F(\BV x(s),\BV y(s),\BV v'(s),\BV w'(s))=F_0(\BV x(s)-t^*\BV
v'(s),\BV y(s)-t^*\BV w'(s),\BV v'(s),\BV w'(s))\\\exp\left(
\lambda\Hside_{\alpha\sigma}(\|\BV x(s)-\BV y(s)\|-\sigma)-\lambda
\Hside_{\alpha\sigma}(\|\BV x(s)-\BV y(s)-t^*(\BV v'(s)-\BV w'(s))\|-\sigma)
\right).
\end{multline}
Without loss of generality, we can take $t^*$ large enough so that the
second term under the exponent is $\lambda$ (that is, the point back
in time $t^*$ is sufficiently far away from the collision zone), which
gives
\begin{multline}
\exp\left(\lambda(\Hside_{\alpha\sigma}(\|\BV x(s)-\BV
y(s)\|-\sigma)-1)\right) \lambda\deriv{}s\Hside_{\alpha\sigma}(\|\BV x(s)-\BV
y(s)\|-\sigma) =\\ =\deriv{}s \exp\left(\lambda(\Hside_{\alpha\sigma}(\|\BV
x(s)-\BV y(s)\| -\sigma) -1)\right).
\end{multline}
The equation for $F$ thus becomes
\begin{multline}
\deriv Fs=F_0(\BV x(s)-t^* \BV v'(s),\BV y(s)-t^*\BV w'(s),\BV
v'(s),\BV w'(s))\\\deriv{}s
\exp\left(\lambda(\Hside_{\alpha\sigma}(\|\BV x(s)-\BV
y(s)\|-\sigma)-1)\right).
\end{multline}
Assuming that the initial condition $F_0$ is continuous, and sending
$\alpha\to 0$, we find, upon traversing the collision zone outward,
\begin{equation}
F=F_{\|\BV x-\BV y\|<\sigma}+(1-e^{-\lambda})F_0',
\end{equation}
where $F_0'$ denotes the initial value of the solution that is carried
from the outside to the point of the collision zone where our
characteristic exits the overlapped region. Assembling the pieces
together, we finally arrive at the following result in the limit as
$\alpha\to 0$: after the characteristic traverses the overlapped
region completely, the solution is given via
\begin{equation}
F=e^{-\lambda}F_0+(1-e^{-\lambda})F_0'.
\end{equation}
Subsequently, sending $\lambda\to\infty$ leads to $F_0$ being
completely replaced by $F_0'$ once the characteristic fully traverses
the overlapped region $\|\BV x-\BV y\|<\sigma$. This is, obviously,
the same solution as the one of the Liouville problem for hard spheres
\eqref{eq:F_zero}--\eqref{eq:BC}. Thus, as $\alpha\to 0$ and
$\lambda\to\infty$, the solution of
\eqref{eq:dyn_sys_2}--\eqref{eq:ht} approximates the solution of
\eqref{eq:xyt}--\eqref{eq:vwxy} in the sense that, for the same
initial condition, the corresponding probability densities of
\eqref{eq:F_zero}--\eqref{eq:BC} and \eqref{eq:kolmogorov2} converge
to each other on the same characteristic curve.

\subsection{A steady solution for many spheres}

To find some steady solutions of \eqref{eq:kolmogorov} for
$0<\lambda<\infty$ and $0<\alpha\ll 1$, we again use the method of
characteristics -- namely, we treat the advection term of
\eqref{eq:kolmogorov} as the derivative in the direction of $(1,\BV
v_1,\ldots,\BV v_K)$, with the directional parameter denoted as
$s$. With this, the variables $(t,\BV x_1,\ldots,\BV x_K)$ are given
via the straight line
\begin{equation}
\label{eq:straight_line}
(t(s),\BV x_1(s),\ldots,\BV x_K(s))=(0,\BV x_{10},\ldots,\BV x_{K0})+
  s(1,\BV v_1,\ldots,\BV v_K).
\end{equation}
Then, along the
straight line~\eqref{eq:straight_line} we have
\begin{equation}
\deriv{}s F(t(s),\BV x_1(s),\ldots,\BV x_K(s))=(1,\BV v_1,\ldots,\BV
v_K)\cdot\parderiv F{(t,\BV x_1,\ldots,\BV x_K)}.
\end{equation}
The corresponding differential equation along the straight line is
subsequently given via
\begin{multline}
\deriv Fs+\sum_{i=1}^{K-1}\sum_{j=i+1}^K\lambda\delta_{\alpha\sigma}
(\|\BV x_i-\BV x_j\|-\sigma)\frac{\BV x_i-\BV x_j}{\|\BV x_i-\BV x_j
  \|}\cdot(\BV v_j-\BV v_i)\\\left[F(\BV v_i',\BV v_j')\Hside\big(
  (\BV x_i-\BV x_j)\cdot(\BV v_i-\BV v_j)\big)+F\Hside\big((\BV x_i
  -\BV x_j)\cdot(\BV v_j-\BV v_i)\big)\right]=0.
\end{multline}
To compute steady solutions, it suffices to look for solutions with
the property $F(\BV v_i',\BV v_j')=F$, for all index pairs $(i,j)$. In
such a situation, the Heaviside step-functions
in~\eqref{eq:kolmogorov} are multiplied by identical $F$'s, and thus
coalesce into 1. The equation for $F$ on the characteristic
\eqref{eq:straight_line} thus becomes
\begin{equation}
\label{eq:steady_state_equation}
\deriv Fs+F\sum_{i=1}^{K-1}\sum_{j=i+1}^K\lambda\delta_{\alpha\sigma}
(\|\BV x_i(s)-\BV x_j(s)\|-\sigma)\frac{\BV x_i(s)-\BV x_j(s)}{\|\BV
  x_i(s)-\BV x_j(s)\|}\cdot(\BV v_j-\BV v_i)=0.
\end{equation}
The equation above in~\eqref{eq:steady_state_equation} can obviously
be integrated via separation of variables, but before we proceed with
that, let us recall \eqref{eq:ht2} and observe that the coefficient in
front of $F$ in~\eqref{eq:steady_state_equation} is by itself the time
derivative of $\lambda\Hside_{\alpha\sigma}$, for each index pair
$(i,j)$. Then, the separation of variables yields the equation
\begin{equation}
\deriv{}s\ln F=-\lambda\sum_{i=1}^{K-1}\sum_{j=i+1}^K\deriv{}s
\Hside_{\alpha\sigma}(\sigma-\|\BV x_i(s)-\BV x_j(s)\|),
\end{equation}
with the following solution on the straight line
\eqref{eq:straight_line}:
\begin{equation}
F=G(\BV x_{10},\ldots,\BV x_{K0},\BV v_1,\ldots,\BV v_K)\exp\left(
-\lambda\sum_{i=1}^{K-1}\sum_{j=i+1}^K\Hside_{\alpha \sigma}(\sigma-
\|\BV x_i(s)-\BV x_j(s)\|)\right).
\end{equation}
Above, $G$ is an arbitrary function of the starting point $(\BV
x_{10},\ldots,\BV x_{K0})$ and the fixed velocity vector $(\BV
v_1,\ldots,\BV v_K)$.

To substitute the obtained $F$ into \eqref{eq:kolmogorov}, we change
the notations back to the original variables:
\begin{equation}
\label{eq:Ft}
F(t,\BV x_1,\ldots,\BV x_K,\BV v_1,\ldots,\BV v_K)=G(\BV x_1-t\BV v_1,
\ldots,\BV x_K-t\BV v_K,\BV v_1,\ldots,\BV v_K)\bar F_K(\BV x_1,
\ldots,\BV x_K),
\end{equation}
where the subscript refers to the number of spheres in the system,
and we denote
\begin{subequations}
\label{eq:steady_state_solution}
\begin{equation}
\bar F_K(\BV x_1,\ldots,\BV x_K)=\frac 1{Z_KS_K}\exp\left(-\lambda
\sum_{i=1}^{K-1}\sum_{j=i+1}^K\Hside_{\alpha\sigma}(\sigma-\|\BV x_i-
\BV x_j\|)\right),
\end{equation}
\begin{equation}
Z_K=\int \exp\left(-\lambda\sum_{i=1}^{K-1}\sum_{j=i+1}^K\Hside_{
  \alpha\sigma}(\sigma-\|\BV x_i-\BV x_j\|)\right)\dif V_K.
\end{equation}
\end{subequations}
Above, $S_K$ is the area of the sphere of zero momentum and constant
energy for $K$ spheres. Observe that $\bar F_K$ is normalized to 1 and
is a probability density by itself.

Now, in order to satisfy the condition $F(\BV v_i',\BV v_j')=F(\BV
v_i,\BV v_j)$ for all index pairs $(i,j)$, we need to have
\begin{equation}
\label{eq:G}
G(\BV x_i-t\BV v_i',\BV x_j-t\BV v_j',\BV v_i',\BV v_j')=G(\BV x_i
-t\BV v_i,\BV x_j-t\BV v_j,\BV v_i,\BV v_j)
\end{equation}
for all $t$, $\BV x_i$, $\BV x_j$, $\BV v_i$ and $\BV v_j$, with $\BV
v_i'$ and $\BV v_j'$ given via \eqref{eq:vwxy}. Recalling that
switching $(\BV v_i,\BV v_j)\to(\BV v_i',\BV v_j')$ preserves the
momentum and energy in \eqref{eq:mom_en}, we have to express $G$ as a
function of these two quantities, at the same time preserving the form
in~\eqref{eq:G}. Apparently, the only form of $G$ which is consistent
with both conditions is given via
\begin{equation}
G=G\left(\frac 1K\sum_{i=1}^K(\BV x_i-t\BV v_i),\sum_{i=1}^K\BV v_i,
\sum_{i=1}^K\|\BV v_i\|^2\right),
\end{equation}
where the division by $K$ is included for convenience. It is not
difficult to see that $F$ of the form in~\eqref{eq:Ft} with $G$ being
of the form~\eqref{eq:G} turns the forward equation
\eqref{eq:kolmogorov} into an identity, and the condition $F(\BV
v_i',\BV v_j')=F(\BV v_i,\BV v_j)$ is satisfied for all index pairs
$(i,j)$.

At this point we recall that $\BV v_1,\ldots,\BV v_K$ belong to the
set of zero momentum in \eqref{eq:mom_en}. This forces $G$ to lose the
dependence on $t$ and on the second argument, while the third argument
becomes the constant energy $E$:
\begin{equation}
\label{eq:GK}
G=G\left(\frac 1K\sum_{i=1}^K\BV x_i,E\right),
\end{equation}
Thus, we found a family of solutions $F$ with $F(\BV v_i,\BV
v_j)=F(\BV v_i',\BV v_j')$ for all pairs $(i,j)$, which are the steady
states of \eqref{eq:kolmogorov}, uniform on the velocity sphere of
zero momentum and constant energy. Apparently, there are infinitely
many such steady states for a given system of hard spheres. However,
note that $G$ above in \eqref{eq:GK} is a function of the center of
mass of the system of spheres, which is invariant under the dynamics
of \eqref{eq:dyn_sys}.

As it will become important later, observe that there is a family of
initial conditions of \eqref{eq:kolmogorov} which has a uniform
distribution of the center of mass of the spheres. For such a family
of initial conditions, $G=1$, and we arrive at the steady solution
which consists purely of $\bar F_K$ in
\eqref{eq:steady_state_equation}. Qualitatively, $\bar F_K(\BV x_1,
\ldots,\BV x_K)$ behaves as follows:
\begin{itemize}
\item $\bar F_K(\BV x_1,\ldots,\BV x_K)$ is constant outside contact
  zones (that is, the regions in the coordinate space where the
  mollifier $\delta_{\alpha\sigma}(\|\BV x_i-\BV x_j\|-\sigma)>0$),
  and its transitions within contact zones are given via the exponents
  of anti-derivatives of the mollifier in \eqref{eq:mollifier}.
\item Outside contact zones, the following condition holds:
\begin{equation}
  \label{eq:F_steady_state}
  \bar F_{\text{overlap}}=e^{-\lambda(n-1)}\bar
  F_{\text{non-overlap}},
\end{equation}
where $n$ is the number of the simultaneously overlapping spheres at a
given point $(\BV x_1,\ldots,\BV x_K)$.
\end{itemize}
Observe that the preceding discussion implicitly assumes that the
steady state of the form in \eqref{eq:steady_state_solution} and
\eqref{eq:GK} is attracting for a given initial condition. This is
certainly not the case for all initial conditions -- there exist other
steady states for which $F(\BV v_i,\BV v_j)\neq F(\BV v_i',\BV v_j')$;
for example, those which are not supported in the contact zones, such
that the spheres do not interact at all.  However, below we will argue
that the state $\bar F_K$ in \eqref{eq:steady_state_solution} is the
most likely candidate for a ``physical'' (that is, statistically most
common) steady state for~\eqref{eq:kolmogorov}.

\subsection{Entropy inequality}

For the $K$-sphere dynamics in \eqref{eq:kolmogorov}, the
conventional Boltzmann (also Shannon \cite{Shan}) entropy is given via
\begin{equation}
\label{eq:entropy}
\cE(F)=-\int F\ln F\dif V_K\dif S_K.
\end{equation}
Here we show that, while the Boltzmann entropy $\cE$ can both increase
and decrease in time, its appropriate modification, known as the
Kullback--Leibler entropy
\cite{AbrMaj3,AbrMajKle,HavMajAbr,KulLei,MajAbrGro},
\begin{equation}
\label{eq:KL_entropy}
P(F,G\bar F_K)=\int F\ln\left(\frac F{G\bar F_K}\right)\dif V_K \dif
S_K,
\end{equation}
is a nonincreasing function of time. Also, unlike the Boltzmann
entropy, the Kullback--Leibler entropy $P$ is invariant under
arbitrary changes of variables, and is also nonnegative, that is, for
two probability densities $F_1$ and $F_2$,
\begin{equation}
P(F_1,F_2)\geq 0,
\end{equation}
where the equality is achieved only when $F_1=F_2$ (or the difference
between the two has zero volume measure on $V_K\times S_K$).

Let $\psi:\RR\to\RR$ be a differentiable function. Then, multiplying
both sides of \eqref{eq:kolmogorov} by its derivative $\psi'$ with $F$
used for the argument, we obtain
\begin{multline}
\left(\parderiv{}t+\sum_{i=1}^K\BV v_i\cdot\parderiv{}{\BV x_i}
\right)\psi(F)=\lambda\sum_{i=1}^{K-1}\sum_{j=i+1}^K\delta_{\alpha
  \sigma}(\|\BV x_i-\BV x_j\|-\sigma)\frac{\BV x_i-\BV x_j}{\|\BV
  x_i-\BV x_j\|}\cdot(\BV v_i-\BV v_j)\\\psi'(F)\left[F(\BV v_i',\BV
  v_j') \Hside \big((\BV x_i-\BV x_j)\cdot(\BV v_i-\BV v_j)
  \big)+F\Hside\big((\BV x_i-\BV x_j)\cdot(\BV v_j-\BV v_i)\big)
  \right].
\end{multline}
Integrating over $\dif V_K\dif S_K$, we obtain
\begin{multline}
\parderiv{}t\int\psi(F)\dif V_K\dif S_K=\lambda \sum_{i=1}^{K-1}
\sum_{j=i+1}^K\int\delta_{\alpha \sigma}(\|\BV x_i-\BV x_j\|-\sigma)
\frac{\BV x_i-\BV x_j}{\|\BV x_i-\BV x_j\|}\cdot(\BV v_i-\BV v_j)
\\\psi'(F)\left[F(\BV v_i',\BV v_j') \Hside \big((\BV x_i-\BV x_j)
  \cdot(\BV v_i-\BV v_j) \big)+F\Hside\big((\BV x_i-\BV x_j)\cdot(\BV
  v_j-\BV v_i)\big) \right]\dif V_K\dif S_K.
\end{multline}
where we denote $F'=F(\BV v',\BV w')$.  Via \eqref{eq:vwxyn},
\eqref{eq:vwxy_inv} and \eqref{eq:det1}, we write, for any $i$ and
$j$, the term with $F(\BV v_i',\BV v_j')$ in the right-hand side as
\begin{multline}
\int\delta_{\alpha\sigma}(\|\BV x_i-\BV x_j\|-\sigma)\frac{\BV x_i-\BV
  x_j}{\|\BV x_i-\BV x_j\|}\cdot(\BV v_i-\BV v_j)\psi'(F)F(\BV v_i',
\BV v_j')\Hside\big((\BV x_i-\BV x_j)\cdot(\BV v_i-\BV v_j)\big)\dif
V_K\dif S_K=\\=\int\delta_{\alpha\sigma}(\|\BV x_i-\BV x_j\|-\sigma)
\frac{\BV x_i-\BV x_j}{\|\BV x_i-\BV x_j\|}\cdot(\BV v_j'-\BV v_i')
\psi'(F)F(\BV v_i',\BV v_j')\Hside \big((\BV x_i-\BV x_j) \cdot(\BV
v_j'-\BV v_i')\big)\dif V_K\dif S_K=\\=\int\delta_{\alpha\sigma}(\|
\BV x_i-\BV x_j\|-\sigma)\frac{\BV x_i-\BV x_j}{\|\BV x_i-\BV x_j\|}
\cdot(\BV v_j-\BV v_i)\psi'(F(\BV v_i',\BV v_j'))F\Hside\big((\BV
x_i -\BV x_j) \cdot(\BV v_j-\BV v_i)\big)\dif V_K\dif S_K,
\end{multline}
which leads to
\begin{multline}
\parderiv{}t\int\psi(F)\dif V_K\dif S_K=\lambda \sum_{i=1}^{K-1}
\sum_{j=i+1}^K\int\delta_{\alpha \sigma}(\|\BV x_i-\BV x_j\|-\sigma)
\frac{\BV x_i-\BV x_j}{\|\BV x_i-\BV x_j\|}\cdot(\BV v_j-\BV v_i)
\\\Hside\big((\BV x_i-\BV x_j)\cdot(\BV v_j-\BV v_i)\big) \left[
  \psi'(F(\BV v_i',\BV v_j'))-\psi'(F)\right]F\dif V_K\dif S_K.
\end{multline}
For the entropy in \eqref{eq:entropy}, we substitute
\begin{equation}
\label{eq:psi_entropy}
\psi(F)=-F\ln F,\qquad \psi'(F)=-(1+\ln F),
\end{equation}
and arrive at the following equation for $\cE$:
\begin{multline}
\label{eq:entropy_t2}
\parderiv\cE t=\lambda\sum_{i=1}^{K-1}\sum_{j=i+1}^K\int
\delta_{\alpha\sigma}(\|\BV x_i-\BV x_j\|-\sigma)\frac{\BV x_i-\BV
  x_j}{\|\BV x_i-\BV x_j\|}\cdot(\BV v_j-\BV v_i) \\\Hside\big((\BV
x_i-\BV x_j)\cdot(\BV v_j-\BV v_i)\big)F\ln\left(\frac F{F(\BV
  v_i',\BV v_j'))}\right)\dif V_K\dif S_K.
\end{multline}
Next, recalling the inequality
\[x\ln x\geq x-1,\]
observe that, for two probability densities $F_1$ and $F_2$, we have
\begin{equation}
\label{eq:FlogF_inequality}
F_1\ln\left(\frac{F_1}{F_2}\right)=F_2\frac{F_1}{F_2}\ln\left(\frac{
  F_1}{F_2}\right)\geq F_2 \left(\frac{F_1}{F_2}-1\right)=F_1-F_2,
\end{equation}
which, upon substitution into \eqref{eq:entropy_t2}, yields the
following inequality:
\begin{multline}
\parderiv\cE t\geq\lambda\sum_{i=1}^{K-1}\sum_{j=i+1}^K\int
\delta_{\alpha\sigma}(\|\BV x_i-\BV x_j\|-\sigma)\frac{\BV x_i-\BV
  x_j}{\|\BV x_i-\BV x_j\|}\cdot(\BV v_j-\BV v_i) \\\Hside\big((\BV
x_i-\BV x_j)\cdot(\BV v_j-\BV v_i)\big)\left(F-F(\BV v_i',\BV
v_j')\right)\dif V_K\dif S_K.
\end{multline}
For the part with $F(\BV v_i',\BV v_j')$, we observe that, for any $i$
and $j$, we have
\begin{multline}
-\int\delta_{\alpha\sigma}(\|\BV x_i-\BV x_j\|-\sigma)\frac{\BV x_i-
  \BV x_j}{\|\BV x_i-\BV x_j\|}\cdot(\BV v_j-\BV v_i)\Hside\big((\BV
x_i-\BV x_j)\cdot(\BV v_j-\BV v_i)\big)F(\BV v_i',\BV v_j')\dif
V_K\dif S_K=\\=-\int\delta_{\alpha\sigma}(\|\BV x_i-\BV
x_j\|-\sigma)\frac{\BV x_i- \BV x_j}{\|\BV x_i-\BV x_j\|}\cdot(\BV
v_i'-\BV v_j')\Hside\big((\BV x_i-\BV x_j)\cdot(\BV v_i'-\BV
v_j')\big)F(\BV v_i',\BV v_j')\dif V_K\dif
S_K=\\=\int\delta_{\alpha\sigma}(\|\BV x_i-\BV x_j\|-\sigma)\frac{\BV
  x_i- \BV x_j}{\|\BV x_i-\BV x_j\|}\cdot(\BV v_j-\BV
v_i)\Hside\big((\BV x_i-\BV x_j)\cdot(\BV v_i-\BV v_j)\big)F\dif
V_K\dif S_K,
\end{multline}
which allows to coalesce the two Heaviside functions into 1 and
results in
\begin{equation}
\label{eq:entropy_t}  
\parderiv\cE t\geq\lambda\sum_{i=1}^{K-1}\sum_{j=i+1}^K\int\delta_{
  \alpha\sigma}(\|\BV x_i-\BV x_j\|-\sigma)\frac{\BV x_i-\BV x_j}{
  \|\BV x_i-\BV x_j\|}\cdot(\BV v_j-\BV v_i)F\dif V_K\dif S_K.
\end{equation}
The right-hand side of \eqref{eq:entropy_t} above can, in general, be
negative, and, therefore, it is possible for the Boltzmann entropy of
the system to decrease in general. As an example, consider the uniform
initial condition in both coordinates and velocities, which, under the
normalization constraint, maximizes the entropy over all possible
states.  However, this state is not a steady state of
\eqref{eq:kolmogorov}; indeed, the states corresponding to the
overlapped spheres will be ``drained'', the solution will become
non-uniform in coordinates, and the entropy will decrease as a result.

To examine the Kullback--Leibler entropy in \eqref{eq:KL_entropy}, let
us look at the expression
\begin{multline}
\parderiv{}t(F\ln(G\bar F_K))=-\ln(G\bar F_K)\sum_{i=1}^K\BV v_i
\cdot \parderiv F{\BV x_i}+\lambda\sum_{i=1}^{K-1}\sum_{j=i+1}^K
\delta_{\alpha\sigma}(\|\BV x_i-\BV x_j\|-\sigma)\ln(G\bar
F_K)\\\frac{ \BV x_i-\BV x_j}{\|\BV x_i-\BV x_j\|}\cdot(\BV v_i-\BV
v_j)\left[F(\BV v_i',\BV v_j')\Hside\big((\BV x_i-\BV x_j)\cdot(\BV
  v_i-\BV v_j)\big)+F\Hside\big((\BV x_i-\BV x_j)\cdot(\BV v_j-\BV
  v_i)\big)\right],
\end{multline}
where $G\bar F_K$ is the steady state from
\eqref{eq:steady_state_solution} and \eqref{eq:GK}. We rearrange the
terms above as
\begin{multline}
\parderiv{}t(F\ln(G\bar F_K))=-\sum_{i=1}^K\BV v_i\cdot\parderiv{}{
  \BV x_i}\left(F\ln(G\bar F_K)\right)+\frac F{G\bar F_K}\sum_{i=1
}^K\BV v_i\cdot\parderiv{}{\BV x_i}(G\bar F_K)+\\+\lambda\sum_{i=1
}^{K-1}\sum_{j=i+1}^K\delta_{\alpha\sigma}(\|\BV x_i-\BV x_j\|-\sigma)
\frac{\BV x_i-\BV x_j}{\|\BV x_i-\BV x_j\|}\cdot(\BV v_i-\BV v_j)
\\\ln(G\bar F_K)\left[F(\BV v_i',\BV v_j')\Hside\big((\BV x_i-\BV
  x_j)\cdot(\BV v_i-\BV v_j)\big)+F\Hside\big((\BV x_i-\BV x_j)\cdot
  (\BV v_j-\BV v_i)\big)\right].
\end{multline}
Using \eqref{eq:kolmogorov} and the fact that $G\bar F_K$ is a
steady state, we write
\begin{equation}
\sum_{i=1}^K\BV v_i\cdot\parderiv{}{\BV x_i}(G\bar F_K)=\lambda
\sum_{i=1}^{K-1}\sum_{j=i+1}^K\delta_{\alpha\sigma}(\|\BV x_i-\BV x_j
\|-\sigma)\frac{\BV x_i-\BV x_j}{\|\BV x_i-\BV x_j\|}\cdot(\BV v_i-\BV
v_j)G\bar F_K,
\end{equation}
and subsequently obtain
\begin{multline}
\parderiv{}t(F\ln(G\bar F_K))=-\sum_{i=1}^K\BV v_i\cdot\parderiv{}{
  \BV x_i}\left(F\ln(G\bar F_K)\right)+
\lambda\sum_{i=1}^{K-1}\sum_{j=i+1}^K\delta_{\alpha\sigma}(\|\BV x_i-
\BV x_j\|-\sigma)\frac{\BV x_i-\BV x_j}{\|\BV x_i-\BV x_j\|}\cdot\\\cdot(\BV
v_i-\BV v_j)\left(F+\ln(G\bar F_K)\left[F(\BV v_i',\BV
  v_j')\Hside\big( (\BV x_i-\BV x_j)\cdot(\BV v_i-\BV
  v_j)\big)+F\Hside\big((\BV x_i- \BV x_j)\cdot(\BV v_j-\BV
  v_i)\big)\right]\right).
\end{multline}
Upon the integration over $\dif V_K\dif S_K$, the terms with
$\ln(G\bar F_K)$ in the right-hand side above disappear, and we arrive
at
\begin{equation}
\label{eq:FGF_t}
\parderiv{}t\int F\ln(G\bar F_K)\dif V_K\dif S_K=\lambda\sum_{i=1}
^{K-1}\sum_{j=i+1}^K\int\delta_{\alpha\sigma}(\|\BV x_i-\BV x_j\|-
\sigma)\frac{\BV x_i-\BV x_j}{\|\BV x_i-\BV x_j\|}\cdot(\BV v_i-\BV
v_j)F\dif V_K\dif S_K.
\end{equation}
Adding \eqref{eq:FGF_t} to \eqref{eq:entropy_t} and changing the sign
on both sides, we arrive at
\begin{equation}
\label{eq:H_theorem}
\parderiv{}t P(F,G\bar F_K)\leq 0,
\end{equation}
that is, the Kullback--Leibler entropy between a solution of
\eqref{eq:kolmogorov} and the steady state in
\eqref{eq:steady_state_solution} and \eqref{eq:GK} is a nonnegative
nonincreasing function of time. The inequality in~\eqref{eq:H_theorem}
is the analog of Boltzmann's $H$-theorem \cite{CerIllPul,Gra,Gols} for
the Kolmogorov equation in \eqref{eq:kolmogorov}.

\subsection{The steady solution for a system with independently
  distributed initial states}

Recall that the factor $G$ in \eqref{eq:GK} is determined by the
distribution of the center of mass of the spheres, which, in turn, is
defined entirely by the corresponding initial condition for the
forward equation \eqref{eq:kolmogorov}. So far, we did not elaborate
on the choice of the initial conditions, and thus $G$ was presumed to
be largely arbitrary. However, from the perspective of physics, some
of the states are more realistic, while others are notably less so. As
an example of a physically unrealistic state, one can arrange the
spheres initially so they fly in straight lines without
interacting. However, in real-world systems, such as gases and
liquids, these states are not encountered in practice, and instead a
strongly chaotic motion of frequently colliding molecules is observed.

Here we propose the ``most common'' steady state for a system with
large number of spheres, based on a ``physicist's reasoning''. Namely,
in what follows, we assume that the spheres are initially distributed
independently of each other. We compute the distribution of their
center of mass, and, observing that it is invariant under the
dynamics, conclude that the relevant steady state must have the same
distribution of the center of mass. Note that the independence of
distributions for each sphere suggests that the initial measures of
overlapped states may be nonzero, however, in our random gas model, it
is permissible for the spheres to overlap.

Let the distribution $F_0$ of the $K$ spheres in the system at the
initial moment of time be given via the product of independent
identical distributions $f_0$ for each sphere:
\begin{equation}
F_0(\BV x_1,\ldots,\BV x_K,\BV v_1,\ldots,\BV v_K)=\prod_{i=1}^Kf_0
(\BV x_i,\BV v_i).
\end{equation}
Let the position of the center of mass of the system be given via $\BV
y$:
\begin{equation}
\BV y=\frac 1K\sum_{i=1}^K\BV x_i.
\end{equation}
Let us express the first coordinate, $\BV x_1$, through the position
of the center of mass $\BV y$, and the rest of the coordinates:
\begin{equation}
\BV x_1=K\BV y-\sum_{i=2}^K\BV x_i.
\end{equation}
Then, the distribution of $\BV y$ is given via
\begin{multline}
g(\BV y)=K^3\int f_0\left(K\BV y-\sum_{i=2}^K\BV x_i,\BV v_1\right)
\prod_{i=2}^Kf_0(\BV x_i,\BV v_i)\dif\BV x_2\ldots\dif\BV x_K\dif\BV
v_1\ldots\dif\BV v_K=\\=K^3\int f_0^x\left(K\BV y-\sum_{i=2}^K\BV
x_i\right)\prod_{i=2}^K f_0^x(\BV x_i)\dif\BV x_2\ldots\dif\BV x_K,
\end{multline}
where $f_0^x$ denotes the $\BV x$-marginal of $f_0$:
\begin{equation}
f_0^x(\BV x)=\int f_0(\BV x,\BV v)\dif\BV v.
\end{equation}
In what follows, we will assume that $f_0^x$ is nondegenerate, that
is, its support is of the same dimension as that of the whole $\BV
x$-space. Let $\hat f$ denote the characteristic function of $f_0^x$,
\begin{equation}
\label{eq:characteristic_function}
\hat f(\BV k)=\int e^{i\BV k\cdot\BV x}f_0^x(\BV x)\dif\BV x,\qquad
f_0^x(\BV x)=\frac 1{(2\pi)^3}\int e^{-i\BV k\cdot\BV x}\hat f(\BV
k)\dif\BV k.
\end{equation}
Clearly, $\hat f(\BV 0)=1$, since $f_0^x$ is a probability density.
Additionally, due to the fact that $f_0^x$ is nondegenerate, $|\hat
f(\BV k)|<1$ for $\BV k\neq\BV 0$ \cite{Tie}.

Next, let us express $g(\BV y)$ via the characteristic function of
$f_0^x$ from \eqref{eq:characteristic_function}:
\begin{multline}
g(\BV y)=\frac{K^3}{(2\pi)^{3K}}\int\dif\BV k_1\ldots\dif\BV k_K\hat
f(\BV k_1)\ldots\hat f(\BV k_K)e^{-iK\BV k_1\cdot\BV y}\\\int e^{i(\BV
  k_1- \BV k_2) \cdot\BV x_2}\ldots e^{i(\BV k_1-\BV k_K)\cdot\BV x_K}
\dif\BV x_2\ldots\dif\BV x_K.
\end{multline}
Recalling that, in the generalized sense,
\begin{equation}
\int e^{i\BV k\cdot\BV x}\dif\BV x=(2\pi)^3\delta(\BV k),
\end{equation}
we obtain
\begin{equation}
g(\BV y)=\frac{K^3}{(2\pi)^3}\int \left(e^{-i\BV k\cdot\BV y}\hat
f(\BV k)\right)^K\dif\BV k.
\end{equation}
Observing that $|e^{-i\BV k\cdot\BV y}\hat f(\BV k)|<1$ for $\BV
k\neq\BV 0$, we find that, in the limit $K\to\infty$, the integrand
approaches zero for all $\BV k\neq\BV 0$. Thus, as $K\to\infty$,
$g(\BV y)$ loses its dependence on $\BV y$ and becomes the uniform
distribution. We thus conclude that the uniform distribution of the
center of mass of the system of spheres is the most physical one.  The
corresponding most physical steady state is thus given via $G=1$ in
\eqref{eq:GK}, as, under $\bar F_K$, any given state of the spheres is
exactly as likely as its arbitrary parallel coordinate translation,
and thus the center of mass of $\bar F_K$ alone is distributed
uniformly.

Observe that the preceding derivation can be modified so that the
spheres are distributed independently, but not identically (that is,
any $i$-th sphere could be distributed independently, but with its own
distribution $f_{0i}$). If so, for the same reasoning to hold, we need
all corresponding characteristic functions $\hat f_i$ of each $\BV
x$-marginal to be bounded above by the same $h(\BV k)$,
\begin{equation}
|\hat f_i(\BV k)|\leq h(\BV k),\qquad h(\BV 0)=1,\quad h(\BV k\neq\BV
0)<1,
\end{equation}
that is, the spheres should initially be distributed independently and
``similarly'', which is a reasonable requirement from the physical
standpoint.

\subsection{The structure of marginal distributions of the physical
steady state}

As it becomes important below, we need to discuss the structure of the
marginal distributions of the physical steady state $\bar F_K(\BV
x_1,\ldots, \BV x_K)$ in~\eqref{eq:steady_state_solution}, that is,
the functions of the form
\begin{equation}
\label{eq:n_marginal}
\bar F_K^{(n)}(\BV x_1,\ldots,\BV x_n,\BV v_1,\ldots,\BV v_n)=S_{n+1
  \ldots K}(\BV v_1,\ldots,\BV v_n)\int\bar F_K(\BV x_1,\ldots,\BV
x_K)\dif V_{n+1 \ldots K},
\end{equation}
where $S_{n+1\ldots K}(\BV v_1,\ldots,\BV v_n)$ is the surface area of
the subset of the constant energy sphere which corresponds to fixed
velocities $\BV v_1, \ldots,\BV v_n$, and $\dif V_{n+1 \ldots K}$ is
the volume element of the subset of the volume which corresponds to
fixed coordinates $\BV x_1,\ldots,\BV x_n$.

The integral over $\dif V_{n+1\ldots K}$ in the right-hand side of
\eqref{eq:n_marginal} can be expressed in the form
\begin{equation}
\label{eq:V_marginal}
\int\bar F_K(\BV x_1,\ldots,\BV x_K)\dif V_{n+1 \ldots K}=\bar F_n(\BV
x_1,\ldots,\BV x_n)\frac{S_n}{S_K}R_K^{(n)}(\BV x_1,\ldots,\BV x_n).
\end{equation}
Above, we introduce the spatial correlation function $R_K^{(n)}(\BV
x_1,\ldots,\BV x_n)$ of the form
\begin{multline}
\label{eq:R_K_n}
R_K^{(n)}(\BV x_1,\ldots,\BV x_n)=\frac{Z_n}{Z_K}\int\bigg(\prod_{i=n+
  1}^Ke^{-\lambda(\Hside_{\alpha\sigma}(\sigma-\|\BV x_1-\BV x_i\|) +
  \ldots+\Hside_{\alpha\sigma}(\|\BV x_n-\BV x_i\|))} \\\prod_{j=i+1
}^K e^{-\lambda\Hside_{\alpha\sigma}(\sigma-\|\BV x_i-\BV x_j\|)}
\bigg)\dif V_{n+1\ldots K},
\end{multline}
that is, the $R_K^{(n)}$ involves integration over all factors
$e^{-\lambda\Hside_{\alpha\sigma}(\sigma-\|\BV x_i-\BV x_j\|)}$ which
are part of $\bar F_K$, but not $\bar F_n$. Observe that the following
normalization conditions hold concurrently:
\begin{subequations}
\begin{equation}
S_n\int\bar F_n(\BV x_1,\ldots,\BV x_n)\dif V_n=1,
\end{equation}
\begin{equation}
S_n\int\bar F_n(\BV x_1,\ldots,\BV x_n)R_K^{(n)}(\BV x_1,\ldots,\BV
x_n) \dif V_n=S_K\int\bar F_K(\BV x_1,\ldots,\BV x_K)\dif V_K=1,
\end{equation}
\end{subequations}
so if $R_K^{(n)}$ does not vary ``too much'' throughout the domain,
its magnitude should generally be of order 1. In situations where the
overlapped states take a very small part of the total volume (which
implies that the volume of the domain greatly exceeds the total volume
of the spheres in it, the so-called ``dilute gas''), the integrand of
\eqref{eq:R_K_n} equals 1 almost everywhere, and, therefore,
$R_K^{(n)}\approx 1$.

For the reasons which will become clear below, of particular
importance are the special cases with $n=1$ and $n=2$. For $n=1$, we
claim
\begin{equation}
\label{eq:1_marginal}
  \bar F_K^{(1)}(\BV v)=\frac 1V\frac{S_{2\ldots K}(\BV v)}{S_K},
\end{equation}
where $V$ is the volume of the coordinate space of a single sphere,
and the independence on $\BV x$ follows from the fact that $\bar F_K$
depends only on the distances between the spheres, and not on their
absolute positions.

For the special case $n=2$, we claim
\begin{equation}
\label{eq:R_K_2}
R_K^{(2)}(\BV x,\BV y)=R_K^{(2)}(\|\BV x-\BV y\|),
\end{equation}
which follows from the symmetry and isotropy considerations.  Indeed,
observe that, since $\bar F_K$ depends only on distances between
different spheres, $R_K^{(2)}$ can depend neither on the absolute
positions of its two spheres, nor on the direction of their mutual
orientation, which leaves only the distance as a possible dependence.
Taking into account \eqref{eq:steady_state_solution}, for the
two-sphere marginal $\bar F_K^{(2)}$ we, therefore, arrive at
\begin{equation}
\label{eq:2_marginal}
\bar F_K^{(2)}(\BV x,\BV y,\BV v,\BV w)=\frac 1{Z_2}e^{-\lambda
  \Hside_{\alpha\sigma}(\sigma-\|\BV x-\BV y\|)}R_K^{(2)}(\|\BV x-\BV
y\|)\frac{S_{3 \ldots K}(\BV v,\BV w)}{S_K}.
\end{equation}
Combining \eqref{eq:1_marginal} and \eqref{eq:2_marginal}, we can
express
\begin{equation}
\label{eq:2_1_marginal}
\bar F_K^{(2)}(\BV x,\BV y,\BV v,\BV w)=\frac{V^2}{Z_2}\frac{S_KS_{3
    \ldots K}(\BV v,\BV w)}{S_{2\ldots K}(\BV v)S_{2\ldots K}(\BV w)}
e^{-\lambda \Hside_{\alpha\sigma}(\sigma-\|\BV x-\BV y\|)}
R_K^{(2)}(\|\BV x-\BV y\|)\bar F_K^{(1)}(\BV v)\bar F_K^{(1)}(\BV w).
\end{equation}

\subsection{The marginal distributions in the limit of infinitely
  many spheres}

In what follows, we are interested in situations where the number of
spheres is large, and thus we need to to examine the limit of
\eqref{eq:2_1_marginal} as $K\to\infty$. In such a limit, first we
observe that
\begin{equation}
\label{eq:VZ_limit}
\lim_{K\to\infty}\frac{V^2}{Z_2}=1.
\end{equation}
This happens due to the fact that the diameter $\sigma$ must decay as
$K^{-1/3}$, in order for the total volume of the spheres to be bounded
(to fit into the domain without overlapping). Therefore, the volume of
the set of points where the integrand of $Z_2$ in
\eqref{eq:steady_state_solution} equals $e^{-\lambda}$ is proportional
to $K^{-1}$. This means that the ratio above behaves as
$V^2/Z_2=1+O(K^{-1})$, and approaches 1 as $K\to\infty$.

Second, it can be shown via geometric arguments (see
\cite{Abr13,AbrKovMaj,MajAbrGro} and references therein) that, as
$K\to\infty$,
\begin{equation}
\label{eq:S_marginal}
\lim_{K\to\infty}\frac{S_{2\ldots K}(\BV v)}{S_K}=\frac{e^{-\|\BV
    v\|^2 /(2\theta)}}{(2\pi\theta )^{3/2}},\qquad
\lim_{K\to\infty}\frac{S_{3\ldots K}(\BV v,\BV w)}{S_K}=\frac
    {e^{-(\|\BV v\|^2+\|\BV w\|^2)/(2\theta)}}{(2\pi\theta)^3},
\end{equation}
where $\theta$ is the temperature of the system of the spheres, given
via
\begin{equation}
E=\frac 32 K\theta.
\end{equation}
The relations in \eqref{eq:S_marginal} lead to
\begin{equation}
\label{eq:S_limit}
\lim_{K\to\infty}\frac{S_KS_{3 \ldots K}(\BV v,\BV w)}{S_{2\ldots K}
  (\BV v)S_{2\ldots K}(\BV w)}=1.
\end{equation}
Combining \eqref{eq:2_1_marginal}, \eqref{eq:VZ_limit} and
\eqref{eq:S_limit}, we arrive, as $K\to\infty$, to
\begin{equation}
\label{eq:F_2_1}
\bar F^{(2)}(\BV x,\BV y,\BV v,\BV w)=e^{-\lambda\Hside_{\alpha\sigma}
  (\sigma-\|\BV x-\BV y\|)}R^{(2)}(\|\BV x-\BV y\|)\bar F^{(1)}(\BV v)
\bar F^{(1)}(\BV w),
\end{equation}
where we dropped the subscript $K$ from $\bar F^{(1)}$, $\bar F^{(2)}$
and $R^{(2)}$ to signify that these quantities are related to
infinitely many spheres.

\section{The forward equation for the marginal distribution of a single
sphere}
\label{sec:forward}

Let us consider an unsteady solution $F$ of the forward equation
in~\eqref{eq:kolmogorov} which is invariant under the permutations of
the spheres, and let us denote the two-sphere and one-sphere marginal
distributions of $F$ as
\begin{equation}
F^{(2)}(t,\BV x,\BV y,\BV v,\BV w)=\int F\dif V_{3\ldots K}\dif
S_{3\ldots K},\qquad f(t,\BV x,\BV v)=\int F^{(2)}\dif\BV y\dif\BV w,
\end{equation}
where the integral in $\BV w$ occurs over the remaining area of $S_K$
for a fixed $\BV v$. Then, integrating~\eqref{eq:kolmogorov} over all
coordinate-velocity pairs $(\BV x_i,\BV v_i)$ but one, we arrive at
\begin{multline}
\label{eq:BBGKY_z}
\parderiv ft+\BV v\cdot\parderiv f{\BV x}=(K-1)\lambda\int\delta_{
  \alpha\sigma}(\|\BV x-\BV y\|-\sigma)\frac{\BV x-\BV y}{\|\BV x-\BV
  y\|}\cdot(\BV v-\BV w)\\\Big[F^{(2)}(\BV x,\BV y,\BV v',\BV w')
  \Hside\big( (\BV x-\BV y)\cdot(\BV v-\BV w)\big)+ F^{(2)}(\BV x,
  \BV y,\BV v,\BV w)\Hside\big((\BV x-\BV y)\cdot(\BV w-\BV v)\big)
  \Big]\dif\BV y\dif\BV w.
\end{multline}
To arrive at the above result in the collision term, observe that, for
fixed $\BV x$ and $\BV y$, we, with the help of~\eqref{eq:vwxyn}
and~\eqref{eq:det1} have, for $i,j>1$, $i\neq j$:
\begin{multline}
\int F(\BV v_i,\BV v_j)\psi(\BV v_i',\BV v_j')\frac{\BV x_i-\BV x_j}{
  \|\BV x_i-\BV x_j\|}\cdot(\BV v_j-\BV v_i)\Hside\big((\BV x_i-\BV
x_j)\cdot(\BV v_j-\BV v_i)\big)\dif S_{2\ldots K}=\\=\int F(\BV v_i,\BV v_j)
\psi(\BV v_i',\BV v_j')\frac{\BV x_i-\BV x_j}{\|\BV x_i-\BV x_j\|}
\cdot(\BV v_i'-\BV v_j')\Hside\big((\BV x_i-\BV x_j)\cdot(\BV v_i'
-\BV v_j')\big)\dif S_{2\ldots K}=\\=-\int F(\BV v_i',\BV v_j')\psi(\BV v_i,\BV
v_j)\frac{\BV x_i-\BV x_j}{\|\BV x_i-\BV x_j\|}\cdot(\BV v_j-\BV
v_i)\Hside\big((\BV x_i-\BV x_j)\cdot(\BV v_i-\BV v_j)\big)\dif S_{2\ldots K},
\end{multline}
which cancels out all terms in the collision which do not involve that
particular sphere over which the integration does not occur. Next,
we switch to the spherical coordinate system and replace
\begin{equation}
\BV y=\BV x\pm\sigma r\BV n,\qquad\dif\BV y=\sigma^3r^2\dif r\dif\BV
n,
\end{equation}
where $r$ is the nondimensional distance, $\BV n$ is the unit vector,
$\dif\BV n$ is the area element of the unit sphere, ``$+$'' is used in
the term with $F^{(2)}(\BV v',\BV w')$, while ``$-$'' is used in the
term with $F^{(2)}(\BV v,\BV w)$. In the new variables, the identity
in \eqref{eq:vwxy} becomes \eqref{eq:vwn}, and \eqref{eq:BBGKY_z}
becomes the leading order identity of the corresponding BBGKY
hierarchy \cite{Bog,BorGre,Kir} for \eqref{eq:kolmogorov}:
\begin{multline}
\label{eq:BBGKY}
\parderiv ft+\BV v\cdot\parderiv f{\BV x}=(K-1)\lambda\sigma^2\int
\delta_\alpha(r-1)\BV n\cdot(\BV w-\BV v)\Hside\big(\BV n\cdot(\BV w-
\BV v)\big)\\\left[F^{(2)}(\BV x,\BV x+\sigma r\BV n,\BV v',\BV w')
  -F^{(2)}(\BV x, \BV x-\sigma r\BV n,\BV v,\BV w)\right]r^2\dif r
\dif\BV n\dif \BV w.
\end{multline}
At this point, it might be tempting to assume that $\alpha$ is small
enough so that one can neglect the variations in $F^{(2)}$ within the
contact zone and integrate in $\dif r$, however, it is clearly not the
case due to e.g. \eqref{eq:F_2_1}. Thus, in order to proceed further,
we first need to find an appropriate closure to $F^{(2)}$ above in
\eqref{eq:BBGKY} in terms of the one-sphere marginal $f$.

\subsection{Approximating the two-sphere marginal via one-sphere
marginals}

Above in~\eqref{eq:BBGKY} we arrived at the ubiquitous closure problem
of molecular dynamics -- the two-sphere marginal $F^{(2)}$ must be
approximated via the one-sphere marginal $f$ in order to be able to
transform the leading order BBGKY identity in \eqref{eq:BBGKY} into a
standalone closed equation for $f$. In the mean field approximation
\cite{Vla} for long-range potentials, the joint density of two
particles is approximated via the product of single-particle
distributions.  However, in our situation, the relation in
\eqref{eq:F_2_1} for the steady state marginals does not permit such a
direct factorization.

Instead, here we choose to relate the unsteady marginals $f$ and
$F^{(2)}$ in the same way their steady counterparts are related in
\eqref{eq:F_2_1} in the limit of infinitely many spheres:
\begin{equation}
\label{eq:F2_approx}
F^{(2)}(t,\BV x,\BV y,\BV w,\BV v)\approx e^{-\lambda\Hside_{\alpha
    \sigma} (\sigma-\|\BV x-\BV y \|)} R^{(2)}(\|\BV x-\BV y\|)
f(t,\BV x,\BV v)f(t,\BV y,\BV w),
\end{equation}
where $R^{(2)}$ depends parametrically on $\alpha$ and $\lambda$:
$R^{(2)}=R_{\lambda,\alpha}^{(2)}$. Observe that the closure in
\eqref{eq:F2_approx} becomes the exact solution if $F^{(2)}$ is the
marginal distribution of the steady state as $K\to\infty$. If
$F^{(2)}$ is not a marginal distribution of the steady state, but
sufficiently close to it, then we can estimate the improvement of
\eqref{eq:F2_approx} over \eqref{eq:F_2_1} via the following
appropriately generalized version of the marginal formula for the
Kullback--Leibler divergence \cite{MajKleCai,AbrMaj}.

Let $F(\BV z_1,\ldots,\BV z_K)$ be a probability density, $\psi(\BV
z_1,\ldots,\BV z_K)>0$ be an integrable function, and $p_1(\BV z_1),
\ldots,p_K(\BV z_K)$ be a set of probability densities for $\BV
z_i\in\RR^d$, $d>0$. Then, the following identity holds for the
Kullback--Leibler divergence $P$:
\begin{equation}
\label{eq:MKC_gen}
P\left(F,\psi\prod_{i=1}^Kp_i(\BV z_i)\right)=P\left(F,\psi
\prod_{i=1}^K f_i(\BV z_i)\right)+\sum_{i=1}^N P(f_i,p_i),
\end{equation}
where $f_i(\BV z_i)$ is the $i$-th marginal of $F$. Indeed, observe
that
\begin{multline}
P\left(F,\psi\prod_{i=1}^Kp_i(\BV z_i)\right)=\int F\left[\ln F-\ln
  \psi- \ln \left(\prod_{i=1}^Kp_i(\BV z_i)\right)\right]\dif\BV Z=\\=
\int F\left[\ln F-\ln\left(\psi\prod_{i=1}^Kf_i(\BV z_i)\right)+\ln
  \left(\prod_{i=1}^K f_i(\BV z_i)\right)-\ln\left(\prod_{i=1}^K
  p_i(\BV z_i) \right) \right]\dif\BV Z,
\end{multline}
whereupon \eqref{eq:MKC_gen} follows from the first and second pair of
terms in the integrand.

The second term in the right-hand side of \eqref{eq:MKC_gen} is always
nonnegative, which bounds the left-hand side below as:
\begin{equation}
\label{eq:MKC_bound}
P\left(F,\psi\prod_{i=1}^Kp_i(\BV z_i)\right)\geq
P\left(F,\psi\prod_{i=1}^K f_i(\BV z_i)\right).
\end{equation}
We can apply the formula above to our set-up by setting $\BV z_1=(\BV
x,\BV v)$, $\BV z_2=(\BV y,\BV w)$, $F=F^{(2)}$, $\psi=e^{-\lambda
  \Hside_{\alpha \sigma}(\sigma-\|\BV x-\BV y \|)} R^{(2)}(\|\BV x-\BV
y\|)$, and $p_1=p_2=\bar F^{(1)}$ from \eqref{eq:F_2_1}. One should
note, however, that the right-hand side of \eqref{eq:MKC_bound} is not
necessarily the Kullback--Leibler divergence due to the fact that
\eqref{eq:F2_approx} (which is the second argument of $P$) is not
necessarily normalized to 1 for large $\sigma$. However, for a dilute
gas (that is, when $\sigma$ is much smaller than the average distance
between the spheres), the second argument of $P$ becomes normalized to
1, so that the right-hand side of the Kullback--Leibler bound in
\eqref{eq:MKC_bound} indeed involves the probability densities in such
a case. Additionally, the closure in \eqref{eq:F2_approx} preserves
the proportionality relations between the velocity moments in $\BV v$
or $\BV w$ variables separately.

Now, taking into account that $\|\BV x-\BV y\|=\sigma r$ above
in~\eqref{eq:BBGKY}, we have
\begin{equation}
\Hside_{\alpha\sigma}(\sigma-\|\BV x-\BV y\|)=\Hside_{\alpha\sigma}(
\sigma-\sigma r)=\Hside_\alpha(1-r),
\end{equation}
and, with the approximation in~\eqref{eq:F2_approx}, the forward
equation for the marginals in~\eqref{eq:BBGKY} becomes
\begin{multline}
\label{eq:forward}
\parderiv ft+\BV v\cdot\parderiv f{\BV x}=(K-1)\lambda\sigma^2\int R_{
  \lambda,\alpha}^{(2)}(\sigma r)e^{-\lambda\Hside_\alpha(1-r)}
\delta_\alpha (r-1)\BV n\cdot(\BV w-\BV v)\Hside\big(\BV n\cdot(\BV w-
\BV v)\big)\\\left[f(\BV x,\BV v')f(\BV x+\sigma r\BV n,\BV w')-f(\BV
  x,\BV v) f(\BV x-\sigma r\BV n,\BV w)\right]r^2 \dif r\dif\BV n
\dif\BV w.
\end{multline}

\subsection{Thin contact zone and impenetrable spheres}

Above in~\eqref{eq:forward}, we can formally assume that the contact
zone is ``thin'', that is, $\alpha\to 0$ so that, for the values of
$r$ for which $\delta_\alpha(r-1)>0$, we have $f(\BV x\pm\sigma r\BV
n)\to f(\BV x\pm\sigma\BV n)$, $R_{\lambda,\alpha}^{(2)}(\sigma r)\to
R_{\lambda,0}^{(2)} (\sigma)$. In such a case, the integral over $\dif
r$ involves only the mollifier $\delta_\alpha(r-1)$ and its
antiderivative, and thus can be integrated across the thin contact
zone exactly:
\begin{equation}
\lambda\int e^{-\lambda\Hside_\alpha(1-r)}\delta_\alpha(r-1)\dif r=
1-e^{-\lambda}.
\end{equation}
This simplifies the forward equation in~\eqref{eq:forward} to
\begin{multline}
\parderiv ft+\BV v\cdot\parderiv f{\BV x}=(K-1)\sigma^2 \left(1-
e^{-\lambda}\right)R_{\lambda,0}^{(2)}(\sigma)\int\BV n\cdot(\BV w-\BV
v)\Hside\big(\BV n\cdot(\BV w-\BV v)\big)\\\left[f(\BV x,\BV v')f(\BV
  x+\sigma\BV n,\BV w')-f(\BV x,\BV v)f(\BV x-\sigma\BV n,\BV w)
  \right]\dif\BV n\dif\BV w.
\end{multline}
Now, observe that the factor $(1-e^{-\lambda})$ in front of the
equation above is the probability that at least one jump arrives in
the point process during the collision. Sending the intensity of the
point process $\lambda\to\infty$ sets this probability to 1, which
means that the spheres can be considered impenetrable. The resulting
forward equation for a single impenetrable sphere is given via
\begin{multline}
\label{eq:enskog}
\parderiv ft+\BV v\cdot\parderiv f{\BV x}=(K-1)\sigma^2 R_{\infty,
  0}^{(2)}(\sigma)\int\BV n\cdot(\BV w-\BV v)\Hside\big(\BV n\cdot
(\BV w-\BV v)\big)\\\left[f(\BV x,\BV v')f(\BV x+\sigma\BV n,\BV w')
  -f(\BV x,\BV v)f(\BV x-\sigma\BV n,\BV w)\right]\dif\BV n\dif\BV w.
\end{multline}
This is a variant of the Enskog equation for hard spheres
\cite{BelLac,Ens,GapGer,Lac,vBeiErn}, which we henceforth refer to as
``the Enskog equation'', even though it is not identical to what was
originally proposed by Enskog. Apparently, the collision integral
closure above in the Enskog equation \eqref{eq:enskog} becomes exact
if $f$ is the single-sphere marginal distribution of the steady state
$\bar F_K$ in~\eqref{eq:steady_state_solution}, in the limit as the
number of spheres $K\to\infty$, the width of the mollifier $\alpha\to
0$, and $\lambda\to\infty$.

Observe that if the gas is dilute (that is, the diameter $\sigma$ of a
sphere is much smaller than the average distance between the spheres),
then we can neglect the higher order terms in $\sigma$ in comparison
to the leading order. This sets $R_{\infty,0}^{(2)}(\sigma)=1$, $f(\BV
x\pm\sigma\BV n,\BV v)=f(\BV x,\BV v)$, and the Enskog equation in
\eqref{eq:enskog} becomes the Boltzmann equation in
\eqref{eq:boltzmann}.

\section{The fluid dynamics of the Enskog equation in a physical
hydrodynamic limit}
\label{sec:limit}

Above, we introduced a new, random model for hard sphere collision
which avoids Loschmidt's objection \cite{Los}, and where the form of
the collision integral is a property of the forward Kolmogorov
equation of the full multisphere dynamics. We have shown that, under
suitable assumptions, this new model leads to the Boltzmann equation
in \eqref{eq:boltzmann}.  However, one of the key differences between
our derivation of the Boltzmann equation and its conventional
derivation \cite{CerIllPul,GalRayTex} is that, as a precursor to the
Boltzmann equation, we also obtain the Enskog equation
\eqref{eq:enskog} in a systematic fashion, by reducing and simplifying
the forward Kolmogorov equation of the multisphere dynamics under
suitable assumptions.

The hydrodynamic limit of the Boltzmann equation is a well-researched
area (see, for example, \cite{Gols} and references therein), and it is
known that the equations for the velocity moments of the Boltzmann
equation lead to the Euler \cite{Bat} and Navier--Stokes
\cite{Gra,ChaCow} equations. However, the hydrodynamic limit of the
Enskog equation is a less researched topic, although there are works
on this subject as well (see \cite{Lac} and references therein).

Below, we derive the fluid dynamics equations of the Enskog equation
in~\eqref{eq:enskog} in the hydrodynamic limit (that is, as the
diameter $\sigma\to 0$) which, on one hand, is physically plausible,
and, on the other hand, produces additional nonvanishing terms in the
resulting fluid dynamics equations.

\subsection{The mass-weighted equation and the hydrodynamic limit}

Here we endow each sphere with a mass density $\rho_{sp}$, which is a
constant parameter. Above, $f$ is the distribution density of a single
sphere in the $K$-sphere system, and thus is normalized to 1. However,
for the subsequent derivation of the fluid dynamics equations, it is
more convenient to normalize the density by the total mass of the
system. With this, we define the mass density $\Mf$ via
\begin{equation}
\Mf(t,\BV x,\BV v)=\frac 16K\pi\rho_{sp}\sigma^3f(t,\BV x,\BV v),
\end{equation}
where the factor in front of $f$ above is the total mass of the system
of $K$ spheres, each of diameter $\sigma$ and density $\rho_{sp}$. The
corresponding mass-weighted form of the Enskog equation for $K$
spheres in \eqref{eq:enskog} is, therefore,
\begin{multline}
\parderiv\Mf t+\BV v\cdot\parderiv\Mf{\BV x}=\frac{K-1}K\frac{6
  R_{\infty,0}^{(2)}(\sigma)}{\pi \rho_{sp}\sigma}\int\BV n\cdot(\BV w
-\BV v)\Hside\big(\BV n\cdot (\BV w-\BV v)\big)\\\left[\Mf(\BV x,\BV
  v') \Mf(\BV x+\sigma\BV n,\BV w')-\Mf(\BV x,\BV v)\Mf(\BV x-\sigma
  \BV n,\BV w)\right]\dif\BV n\dif\BV w.
\end{multline}
Here, observe that, as $K\to\infty$, the ratio $(K-1)/K\to 1$, and we
need to decide how the remaining coefficient in front of the integral
behaves as $\sigma\to 0$. Here, we argue that the density $\rho_{sp}$
of the spheres should be kept constant;
\begin{figure}
\includegraphics[width=0.5\textwidth]{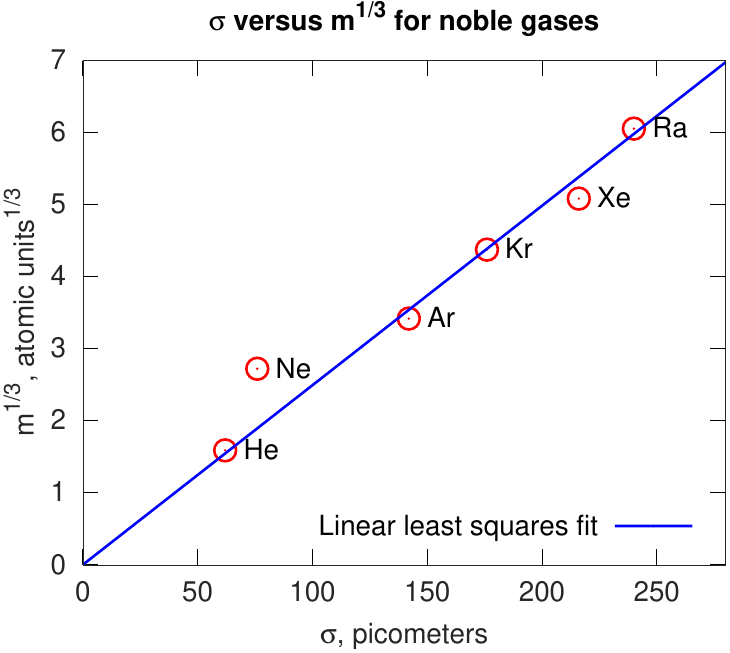}
\caption{Atomic diameter vs atomic mass for noble gases. The values of
  the atomic diameter are taken from~\cite{CleRaiRei}. The values of
  the atomic mass are taken from the standard periodic table of
  elements.}
\label{fig:sigma_vs_m}
\end{figure}
in Figure~\ref{fig:sigma_vs_m}, we plot the cubic root of the atomic
mass versus the atomic diameter for noble gases, so that the straight
line of the least squares fit approximates the hypothetical situation
with constant density of the spheres $\rho_{sp}$.  Observe that, with
the exception of neon, all atoms align on the least squares fit line,
which indicates that the density of atoms of the actual noble gases
tends to be constant. The resulting Enskog equation is given via
\begin{multline}
\label{eq:enskog_constant_density}
\parderiv\Mf t+\BV v\cdot\parderiv\Mf{\BV x}=\frac{6R_{\infty,0}^{(2)}
  (\sigma)}{\pi\rho_{sp}\sigma}\int\BV n\cdot(\BV w-\BV v)\Hside\big(
\BV n\cdot (\BV w-\BV v)\big)\\\left[\Mf(\BV x,\BV v')\Mf(\BV x+\sigma
  \BV n,\BV w')-\Mf(\BV x,\BV v)\Mf(\BV x-\sigma \BV n,\BV w)\right]
\dif\BV n\dif\BV w,
\end{multline}
where the factor in front of the collision integral above behaves as
$\sim\sigma^{-1}$ in the constant-density hydrodynamic limit
$K\to\infty$, $\sigma\to 0$, $\rho_{sp}=$ const.

\subsection{The Euler equations}

With $\sigma$ being treated as a small parameter, a standard way to
look for a solution of \eqref{eq:enskog_constant_density} is to expand
$\Mf$ in powers of $\sigma$ \cite{Lac}:
\begin{equation}
\label{eq:f_expansion}
\Mf(t,\BV x,\BV v,\sigma)=\Mf_0(t,\BV x,\BV v)+\sigma\Mf_1(t,\BV x,\BV
v)+\sigma^2\Mf_2(t,\BV x,\BV v)+\ldots,
\end{equation}
and then recover the expansion terms sequentially starting from the
leading order. We also note that $R_{\infty,0}^{(2)}(0)=1$, and thus
\begin{equation}
R_{\infty,0}^{(2)}(\sigma)=1+\sigma\deriv{}{\sigma}R_{\infty,0}^{(2)}(\sigma)
\Big|_{\sigma=0}+\frac{\sigma^2}2\deriv{^2}{\sigma^2}R_{\infty,0}^{(2)}(\sigma)
\Big|_{\sigma=0}+\ldots
\end{equation}
As a result, in the leading order of $\sigma$, we find
\begin{equation}
\label{eq:leading_order}
\int\BV n\cdot(\BV w-\BV v)\Hside\big(\BV n\cdot(\BV w-\BV v)\big)
\left[\Mf_0(\BV v')\Mf_0(\BV w')-\Mf_0(\BV v)\Mf_0(\BV w)\right] \dif
\BV n \dif\BV w=0,
\end{equation}
which is the usual Boltzmann collision integral
\cite{CerIllPul,ChaCow,Gra,Gols}.  This means that the leading order
expansion term $\Mf_0$ is the standard Maxwell--Boltzmann
thermodynamic equilibrium state,
\begin{equation}
\label{eq:MB_state}
\Mf_0(t,\BV x,\BV v)=\frac\rho{(2\pi\theta)^{3/2}}
\exp\left(-\frac{\|\BV v-\BV u\|^2}{2\theta}\right),
\end{equation}
where the mass density $\rho(t,\BV x)$, average velocity $\BV u(t,\BV
x)$ and kinetic temperature $\theta(t,\BV x)$ are given by the
following mass-weighted moments of molecular velocity:
\begin{equation}
\label{eq:rho_u_theta}
\rho=\int\Mf_0\dif\BV v,\qquad\rho\BV u=\int\BV v\Mf_0\dif\BV v,\qquad
\rho\left(\frac 12\|\BV u\|^2+\frac 32\theta\right)=\rho\epsilon=\frac
12\int\|\BV v\|^2\Mf_0\dif\BV v.
\end{equation}
Above, for further convenience, we additionally denoted the average
kinetic energy $\epsilon$ via
\begin{equation}
\epsilon=\frac 12\|\BV u\|^2+\frac 32\theta.
\end{equation}
For the next order in $\sigma$, we have
\begin{multline}
\label{eq:next_order}
\parderiv{\Mf_0}t+\BV v\cdot\parderiv{\Mf_0}{\BV x}=\frac 6{\pi
  \rho_{sp}} \int\BV n\cdot(\BV w-\BV v)\Hside\big(\BV n\cdot(\BV w-
\BV v)\big) \bigg[\big(\Mf_0(\BV v')\Mf_1(\BV w')+\Mf_1(\BV v')
  \Mf_0(\BV w')-\\-\Mf_0(\BV v)\Mf_1(\BV w)-\Mf_1(\BV v)\Mf_0(\BV w)
  \big)+\BV n\cdot\left(\Mf_0(\BV v') \parderiv{\Mf_0}{\BV x}(\BV w')
  +\Mf_0(\BV v)\parderiv{\Mf_0}{\BV x}(\BV w) \right)\bigg] \dif\BV
n\dif\BV w.
\end{multline}
Observe that the form of the equation above generally does not
guarantee that $\Mf_0$ retains its Gaussian form
in~\eqref{eq:MB_state}, due to the additive collision integral in the
right-hand side of \eqref{eq:next_order}. Thus, we will have to resort
to a weaker formulation of \eqref{eq:next_order} in the form of
velocity moments of 1, $\BV v$, and $\|\BV v\|^2$, in the same manner
as in \cite{Gra,Lac}, which allows to exclude the terms with $\Mf_1$
completely and close the equation with respect to $\rho$, $\BV u$, and
$\theta$.

First, for two arbitrary functions $\psi(\BV v)$ and $h(\BV v,\BV w)$,
the latter with the symmetry property $h(\BV v,\BV w)=h(\BV w,\BV v)$,
observe the following identities (assuming that the integrals are
bounded):
\begin{subequations}
\label{eq:id1}
\begin{multline}
\int\psi(\BV v)h(\BV v,\BV w)\BV n\cdot(\BV w-\BV v)\Hside\big(\BV
n\cdot(\BV w-\BV v)\big)\dif\BV n\dif\BV w\dif\BV v=\\=\int\psi(\BV
w)h(\BV v,\BV w)\BV n\cdot(\BV w-\BV v)\Hside\big(\BV n\cdot(\BV w-\BV
v)\big)\dif\BV n\dif\BV w\dif\BV v,
\end{multline}
\begin{multline}
\int\psi(\BV v')h(\BV v,\BV w)\BV n\cdot(\BV w-\BV v)\Hside\big(\BV
n\cdot(\BV w-\BV v)\big)\dif\BV n\dif\BV w\dif\BV v=\\=\int\psi(\BV
w')h(\BV v,\BV w)\BV n\cdot(\BV w-\BV v)\Hside\big(\BV n\cdot(\BV
w-\BV v)\big)\dif\BV n\dif\BV w\dif\BV v.
\end{multline}
\end{subequations}
Above, with help of the symmetry of $h(\BV v,\BV w)$, we renamed $\BV
v$ as $\BV w$ and vice versa, and also changed the sign of $\BV n$.

Second, observe the following chain of identities:
\begin{multline}
\label{eq:id2}
\int\psi(\BV v)h(\BV v',\BV w')\BV n\cdot(\BV w-\BV v)\Hside\big(\BV
n\cdot(\BV w-\BV v)\big)\dif\BV n\dif\BV w\dif\BV v=\\=\int\psi(\BV
v)h(\BV v',\BV w')\BV n\cdot(\BV v'-\BV w')\Hside\big(\BV n\cdot(\BV
v'-\BV w')\big)\dif\BV n\dif\BV w\dif\BV v=\\=\int\psi(\BV v)h(\BV
v',\BV w')\BV n\cdot(\BV w'-\BV v')\Hside\big(\BV n\cdot(\BV w'-\BV
v') \big)\dif\BV n\dif\BV w'\dif\BV v'=\\=\int\psi(\BV v')h(\BV v,\BV
w)\BV n\cdot(\BV w-\BV v)\Hside\big(\BV n\cdot(\BV w-\BV v)\big)
\dif\BV n\dif\BV w\dif\BV v.
\end{multline}
Above, in the first identity we used \eqref{eq:vwxyn}; in the second
identity we used \eqref{eq:det1} and also replaced $\BV n$ with $-\BV
n$, using the fact that $\BV v'$ and $\BV w'$ in \eqref{eq:vwn} are
invariant with respect to the sign of $\BV n$; in the last identity we
renamed $\BV v'$ as $\BV v$, $\BV w'$ as $\BV w$, and vice versa.

Third, with help of \eqref{eq:id1} and \eqref{eq:id2}, we can write
\begin{multline}
\label{eq:id3}
\int\psi(\BV v)\big(h(\BV v',\BV w')-h(\BV v,\BV w)\big)\BV n\cdot(\BV
w-\BV v)\Hside\big(\BV n\cdot(\BV w-\BV v)\big)\dif\BV n\dif\BV w\dif
\BV v=\\=\int\big(\psi(\BV v')-\psi(\BV v)\big)h(\BV v,\BV w)\BV n
\cdot(\BV w-\BV v)\Hside\big(\BV n\cdot(\BV w-\BV v)\big)\dif\BV n\dif
\BV w\dif\BV v=\\=\frac 12\int\big(\psi(\BV v')+\psi(\BV w')-\psi(\BV
v)-\psi(\BV w)\big)h(\BV v,\BV w)\BV n \cdot(\BV w-\BV v) \Hside
\big(\BV n\cdot(\BV w-\BV v)\big)\dif\BV n\dif \BV w\dif\BV v.
\end{multline}
For a constant $\psi$, the integral above is, obviously, zero
regardless of the choice of $h(\BV v,\BV w)$. Additionally, recall
that $\BV v'$ and $\BV w'$ in \eqref{eq:vwn} are chosen so that the
momentum and energy of any two spheres are preserved during the
collision. Thus, if $\psi(\BV v)$ is set to $\BV v$ or $\|\BV v\|^2$
(or any linear combination thereof, including an additive constant),
then the integral above in \eqref{eq:id3} is also zero irrespective of
the choice of $h(\BV v,\BV w)$.

Computing the velocity moments in \eqref{eq:rho_u_theta} of both sides
of the equation in \eqref{eq:next_order}, and setting $h(\BV v,\BV
w)=\Mf_0(\BV v)\Mf_1(\BV w)+\Mf_1(\BV v)\Mf_0(\BV w)$, we arrive at
\begin{subequations}
\label{eq:transport}
\begin{equation}
\parderiv\rho t+\parderiv{}{\BV x}\cdot(\rho\BV u)=0,\qquad
\parderiv{(\rho\BV u)}t+\parderiv{}{\BV x}\cdot(\rho(\BV u\BV u^T+
\theta\BM I))= \coll[\BV v],
\end{equation}
\begin{equation}
\parderiv{(\rho\epsilon)}t+\parderiv{}{\BV x}\cdot \left(\rho
(\epsilon+\theta)\BV u\right)=\frac 12\coll[\|\BV v\|^2],
\end{equation}  
\end{subequations}
where by $\coll$ we denote the corresponding collision integrals
\begin{subequations}
\begin{equation}
\coll[\BV v]=\frac 6{\pi\rho_{sp}}\int(\BV v-\BV v')\BV n\cdot(\BV w-
\BV v)\Hside \big(\BV n\cdot(\BV w-\BV v)\big)\BV n\cdot\parderiv{
  \Mf_0(\BV w)}{\BV x}\Mf_0(\BV v)\dif\BV n\dif\BV w\dif\BV v,
\end{equation}
\begin{multline}
\coll[\|\BV v\|^2]=\frac 6{\pi\rho_{sp}}\int\left(\|\BV v\|^2-\|\BV v'
\|^2\right)\BV n\cdot(\BV w-\BV v)\\\Hside\big(\BV n\cdot(\BV w-\BV v)
\big)\BV n\cdot\parderiv{\Mf_0(\BV w)}{\BV x}\Mf_0(\BV v)\dif\BV n
\dif\BV w\dif\BV v.
\end{multline}
\end{subequations}
Next, recalling from~\eqref{eq:vwn} that
\begin{equation}
\BV v-\BV v'=-\big(\BV n\cdot(\BV w-\BV v)\big)\BV n,\qquad\|\BV v\|^2
-\|\BV v'\|^2=-\big(\BV n\cdot(\BV w+\BV v)\big)\big(\BV n\cdot(\BV w
-\BV v) \big),
\end{equation}
we further obtain
\begin{subequations}
\begin{equation}
\coll[\BV v]=-\frac 6{\pi\rho_{sp}}\int\big(\BV n\cdot(\BV w-\BV v)
\big)^2\Hside\big(\BV n \cdot(\BV w-\BV v)\big) \BV n\cdot\parderiv{
  \Mf_0(\BV w)}{\BV x}\Mf_0(\BV v)\BV n\dif\BV n\dif\BV w\dif\BV v,
\end{equation}
\begin{multline}
\coll[\|\BV v\|^2]=-\frac 6{\pi\rho_{sp}}\int\big(\BV n\cdot(\BV w+\BV
v)\big)\big(\BV n\cdot(\BV w-\BV v)\big)^2\\\Hside\big(\BV n\cdot(\BV
w-\BV v)\big)\BV n\cdot\parderiv{\Mf_0(\BV w)}{\BV x}\Mf_0(\BV v) \dif
\BV n\dif\BV w\dif\BV v.
\end{multline}  
\end{subequations}
With $\Mf_0$ provided explicitly in \eqref{eq:MB_state}, the
expressions above are integrated exactly into
\begin{equation}
\label{eq:euler_collision}
\coll[\BV v]=-\frac 4{\rho_{sp}}\parderiv{}{\BV
  x}\left(\rho^2\theta\right),\qquad
\coll[\|\BV v\|^2]=-\frac 8{\rho_{sp}}\parderiv{}{\BV
  x}\cdot\left(\rho^2\theta\BV u\right).
\end{equation}  
Substituting \eqref{eq:euler_collision} into \eqref{eq:transport}, we
arrive at
\begin{subequations}
\label{eq:enskog_euler}
\begin{equation}
\parderiv\rho t+\parderiv{}{\BV x}\cdot(\rho\BV u)=0,\qquad
\parderiv{(\rho\BV u)}t+\parderiv{}{\BV x}\cdot\left(\rho\left(\BV u
\BV u^T+\left(1+\frac{4\rho}{\rho_{sp}}\right)\theta\BM I\right)
\right)=0,
\end{equation}
\begin{equation}
\parderiv{(\rho\epsilon)}t+\parderiv{}{\BV x}\cdot\left(\rho\left(
\epsilon+\left(1+\frac{4\rho}{\rho_{sp}}\right)\theta\right)\BV u
\right)=0.
\end{equation}  
\end{subequations}
The equations above were also obtained in \cite{Lac}, where they were
referred to as the ``Enskog--Euler'' equations. With
$\rho/\rho_{sp}\ll 1$ (dilute gas), the equations in
\eqref{eq:enskog_euler} become the conventional Euler equations
\cite{Bat,Gols}. However, observe that in the physical
constant-density limit, considered here, the additional terms are
formally nonvanishing despite the fact that the diameter of the sphere
$\sigma\to 0$.

\subsection{The Newton and Fourier laws}

Here we examine the next-order term $\Mf_1$ from
\eqref{eq:f_expansion}.  With $\Mf_0$ already computed above in
\eqref{eq:MB_state}, we can use \eqref{eq:next_order} to determine the
properties of $\Mf_1$. However, observe that, while the form of
$\Mf_0$ is given uniquely by the leading order identity
\eqref{eq:leading_order}, such a strict constraint on $\Mf_1$ is not
present -- indeed, the next-order equation in \eqref{eq:next_order}
does not call for any particular restriction on $\Mf_1$. Therefore,
the form of $\Mf_1$ in the expansion \eqref{eq:f_expansion} can be
chosen as necessary.

Below, we are going to follow \cite{Gra} and assume that the full
density $\Mf$ can be expressed in the form of the Hermite polynomial
expansion around the Gaussian density $\Mf_0$. This form is chosen
solely due to convenience -- as shown below, all subsequent moment
integrals that arise are expressed in terms of elementary
functions. However, this is not a unique choice for $\Mf$; for an
alternative, see, for example, \cite{Lev}, where the form of density
was chosen so as to maximize the Boltzmann entropy under given moment
constraints. The downside of the latter approach is that the moment
integrals of the solutions of constrained maximum entropy optimization
problems cannot, in general, be expressed in terms of elementary
functions.

Following \cite{Gra}, we choose $\Mf_1$ in the form of the Hermite
polynomial which includes only the following powers of $\BV v$: $\BV
v$ itself, $\BV v\BV v^T$, and $\|\BV v\|^2\BV v$. We impose the
following orthogonality conditions on $\Mf_1$:
\begin{equation}
\int\Mf_1\dif\BV v=0,\qquad\int\BV v\Mf_1\dif\BV v=\BV 0,\qquad
\int\|\BV v\|^2\Mf_1\dif\BV v=0.
\end{equation}
The following higher-order centered moments of $\Mf_1$ will be used as
the new variables in addition to $\rho$, $\BV u$ and $\theta$:
\begin{equation}
\int(\BV v-\BV u)(\BV v-\BV u)^T\Mf_1\dif\BV v=\rho\BM S,\qquad\frac
12 \int\|\BV v-\BV u\|^2(\BV v-\BV u)\Mf_1\dif\BV v=\rho\BV q.
\end{equation}
Above, $\BM S$ and $\BV q$ are called the stress and heat flux,
respectively. Note that the symmetric matrix $\BM S$ has zero trace
due to the orthogonality requirements above. Then, according to the
calculations in \cite{Gra3,Gra}, $\Mf_1$ is given by the following
Hermite polynomial in $\BV v$:
\begin{equation}
\label{eq:f1}
\Mf_1=\frac 1{\theta^2}\Mf_0\left((\BV u-\BV v)\cdot\BV q+\frac 12(\BV
v-\BV u)^T\BM S(\BV v-\BV u)+\frac 1{5\theta}\|\BV v-\BV u\|^2(\BV
v-\BV u)\cdot\BV q\right).
\end{equation}
With the ansatz in \eqref{eq:f1}, we only need to relate the stress
$\BM S$ and heat flux $\BV q$ to the leading order approximation
$\Mf_0$ (that is, to $\rho$, $\BV u$ and $\theta$). To achieve that,
we proceed via integrating \eqref{eq:next_order} against the velocity
moments $\BV v\BV v^T$ and $\|\BV v\|^2\BV v$:
\begin{subequations}
\begin{equation}
\parderiv{}t\int(\BV v\BV v^T)\Mf_0\dif\BV v+\parderiv{}{\BV x} \cdot
\int \BV v(\BV v\BV v^T)\Mf_0\dif\BV v=\coll[\BV v\BV v^T],
\end{equation}
\begin{equation}
\parderiv{}t\int\|\BV v\|^2\BV v\Mf_0\dif\BV v+\parderiv{}{\BV x}\cdot
\int\|\BV v\|^2(\BV v\BV v^T)\Mf_0\dif\BV v=\coll[\|\BV v\|^2\BV v].
\end{equation}
\end{subequations}
Then, we switch to the centered moments:
\begin{subequations}
\begin{equation}
\BV v\BV v^T=(\BV v-\BV u)(\BV v-\BV u)^T+(\BV v-\BV u)\BV u^T+\BV
u(\BV v-\BV u)^T+\BV u\BV u^T,
\end{equation}
\begin{equation}
\|\BV v\|^2\BV v=\|\BV v-\BV u\|^2(\BV v-\BV u)+2(\BV v-\BV u)(\BV
v-\BV u)^T\BV u+\|\BV v\|^2\BV u+\|\BV u\|^2\BV v-\|\BV u\|^2\BV u.
\end{equation}
\end{subequations}
Substituting the above decompositions and using the moment definitions
of $\rho$, $\BV u$ and $\theta$, we arrive at
\begin{subequations}
\begin{multline}
\parderiv{}t(\rho(\BV u\BV u^T+\theta\BM I))+\parderiv{(\rho\theta\BV
  u)}{\BV x}+\parderiv{(\rho\theta\BV u)}{\BV x}^T +\left(\parderiv{
}{\BV x}\cdot(\rho\theta\BV u)\right)\BM I+\parderiv{}{\BV x}\cdot
(\rho\BV u\otimes\BV u\otimes\BV u)=\\=\coll[(\BV v-\BV u)(\BV v-\BV u
  )^T]+\coll[\BV v]\BV u^T +\BV u\coll[\BV v]^T,
\end{multline}
\begin{multline}
2\parderiv{}t\left(\rho(\epsilon+\theta)\BV u\right)+\parderiv{}{\BV
  x}\cdot\left(5\rho\theta^2\BM I+7\rho\theta\BV u\BV
u^T+\rho\theta\|\BV u\|^2\BM I+\rho\|\BV u\|^2\BV u\BV
u^T\right)=\\=\coll[\|\BV v-\BV u\|^2(\BV v-\BV u)]+2\coll[(\BV v-\BV
  u)(\BV v-\BV u)^T]\BV u+\coll[\|\BV v\|^2]\BV u+\|\BV u\|^2\coll[\BV
  v].
\end{multline}
\end{subequations}
We exclude the time derivatives above via the Enskog--Euler equations
in \eqref{eq:enskog_euler}:
\begin{subequations}
\begin{equation}
\coll[(\BV v-\BV u)(\BV v-\BV u)^T]=\rho\theta\left(\parderiv{\BV
  u}{\BV x}+\parderiv{\BV u}{\BV x}^T-\frac 23\left(\parderiv{}{\BV
  x}\cdot\BV u\right)\right)+\frac 13\coll[\|\BV v\|^2]\BM I-\frac
23\BV u^T\coll[\BV v]\BM I,
\end{equation}
\begin{equation}
\coll[\|\BV v-\BV u\|^2(\BV v-\BV u)]=5\rho\theta\parderiv\theta{\BV
  x}+5\theta\coll[\BV v].
\end{equation}
\end{subequations}
Recalling the form of the collision terms, and using \eqref{eq:vwn},
we further obtain
\begin{subequations}
\label{eq:S_q}
\begin{multline}
\frac 6{\pi\rho_{sp}}\int\big(\BV n\cdot(\BV w-\BV
v)\big)^2\left(\big(\BV n\cdot(\BV w-\BV v)\big)\BV n\BV n^T+\BV n(\BV
v-\BV u)^T+(\BV v-\BV u)\BV n^T\right)\Hside\big(\BV n\cdot(\BV w-\BV
v)\big)\\\left(\Mf_0(\BV v)\Mf_1(\BV w)+\Mf_1(\BV v)\Mf_0(\BV w)-\BV n\cdot
  \Mf_0(\BV v) \parderiv{\Mf_0}{\BV x}(\BV w)\right)\dif\BV n\dif\BV
w\dif\BV v=\\= \rho\theta\left(\parderiv{\BV u}{\BV x}+\parderiv{\BV
  u}{\BV x}^T -\frac 23\left(\parderiv{}{\BV x}\cdot\BV u\right)\BM
I\right)-\frac 8{3\rho_{sp}}\rho^2\theta\left(\parderiv{}{\BV
  x}\cdot\BV u\right)\BM I,
\end{multline}
\begin{multline}
\frac 6{\pi\rho_{sp}}\int(\BV n\cdot(\BV w-\BV v))^2\Big[(\BV n\cdot
  (\BV w-\BV v))^2\BV n+(\BV n\cdot(\BV w-\BV v))(\BM I+2\BV n\BV n^T)
  (\BV v-\BV u)+\\+\left(\|\BV v-\BV u\|^2\BM I+2(\BV v-\BV u)(\BV v-
  \BV u)^T\right)\BV n\Big]\Hside\big(\BV n\cdot(\BV w-\BV v)\big)
\\\left(\Mf_0(\BV v)\Mf_1(\BV w)+\Mf_1(\BV v)\Mf_0(\BV w)-\BV n\cdot
\Mf_0(\BV v)\parderiv{\Mf_0}{\BV x}(\BV w)\right)\dif\BV n\dif\BV
w\dif\BV v=5\rho\theta\parderiv\theta{\BV x}-\frac{20}{
  \rho_{sp}}\theta\parderiv{(\rho^2\theta)}{\BV x}.
\end{multline}
\end{subequations}
Taking into account the explicit formulas for $\Mf_0$ in
\eqref{eq:MB_state} and $\Mf_1$ in \eqref{eq:f1}, we obtain the
following exact relations via direct integration:
\begin{subequations}
\label{eq:rel1}
\begin{multline}
\int\big(\BV n\cdot(\BV w-\BV v)\big)^2\left(\big(\BV n\cdot(\BV w-\BV
v)\big)\BV n\BV n^T+\BV n(\BV v-\BV u)^T+(\BV v-\BV u)\BV n^T\right)
\\\Hside\big(\BV n\cdot(\BV w-\BV v)\big)\left(\Mf_0(\BV v) \Mf_1(\BV
w)+\Mf_1(\BV v)\Mf_0(\BV w)\right)\dif\BV n\dif\BV w\dif\BV
v=-\frac{16\sqrt\pi}5\rho^2\sqrt\theta\BM S,
\end{multline}
\begin{multline}
\int\big(\BV n\cdot(\BV w-\BV v)\big)^2\left(\big(\BV n\cdot(\BV w-\BV
v)\big)\BV n\BV n^T+\BV n(\BV v-\BV u)^T+(\BV v-\BV u)\BV n^T\right)
\\\Hside\big(\BV n\cdot(\BV w-\BV v)\big)\BV n\cdot \Mf_0(\BV v)
\parderiv{\Mf_0}{\BV x}(\BV w)\dif\BV n\dif\BV w\dif\BV v=\frac{4\pi}{15
}\rho^2\theta\left(\parderiv{\BV u}{\BV x}+\parderiv{\BV u}{\BV
  x}^T+\left(\parderiv{}{\BV x}\cdot\BV u\right)\BM I\right),
\end{multline}
\begin{multline}
\int(\BV n\cdot(\BV w-\BV v))^2\Big[(\BV n\cdot(\BV w-\BV v))^2\BV n+
  (\BV n\cdot(\BV w-\BV v))(\BM I+2\BV n\BV n^T)(\BV v-\BV u)+\\+
  \left(\|\BV v-\BV u\|^2\BM I+2(\BV v-\BV u)(\BV v-\BV u)^T\right)
  \BV n\Big]\Hside\big(\BV n\cdot(\BV w-\BV v)\big)\\\left(\Mf_0(\BV v)
\Mf_1(\BV w)+\Mf_1(\BV v)\Mf_0(\BV w)\right)\dif\BV n\dif\BV w\dif\BV v=
-\frac{64\sqrt\pi}{15}\rho^2\sqrt\theta\BV q,
\end{multline}
\begin{multline}
\int(\BV n\cdot(\BV w-\BV v))^2\Big[(\BV n\cdot(\BV w-\BV v))^2\BV n+
  (\BV n\cdot(\BV w-\BV v))(\BM I+2\BV n\BV n^T)(\BV v-\BV u)+\\+
  \left(\|\BV v-\BV u\|^2\BM I+2(\BV v-\BV u)(\BV v-\BV u)^T\right)
  \BV n\Big]\Hside\big(\BV n\cdot(\BV w-\BV v)\big) \\\BV n\cdot
\Mf_0(\BV v)\parderiv{\Mf_0}{\BV x}(\BV w)\dif\BV n\dif\BV w\dif\BV v=
\frac{10\pi}3\theta\parderiv{(\rho^2\theta)}{\BV x}+2\pi\rho^2
\theta\parderiv\theta{\BV x}.
\end{multline}
\end{subequations}
Substituting the identities in \eqref{eq:rel1} into \eqref{eq:S_q}, we
express the stress $\BM S$ and heat flux $\BV q$ in terms of $\rho$,
$\BV u$ and $\theta$:
\begin{subequations}
\label{eq:S_q_NS}
\begin{equation}
\rho\BM S=\left(1+\frac{8\rho}{5\rho_{sp}}\right)[\rho\BM
  S]_B=-\left(1+\frac{8\rho}{5\rho_{sp}}\right)\mu\left(\parderiv{ \BV
  u}{\BV x}+\parderiv{\BV u}{\BV x}^T-\frac 23\left(\parderiv{ }{\BV
  x}\cdot\BV u\right)\BM I \right),
\end{equation}
\begin{equation}
\rho\BV q=\left(1+\frac{12\rho}{5 \rho_{sp}}\right)[\rho\BV
  q]_B=-\left(1+\frac{12\rho}{5 \rho_{sp}}\right)\frac{15}4\mu
\parderiv\theta{\BV x},\qquad\mu=\frac{5\sqrt\pi\rho_{sp}\sigma}{
  96}\sqrt\theta,
\end{equation}
\end{subequations}
where $\mu$ is the usual viscosity for the hard sphere gas \cite{Gra},
and $[\rho\BM S]_B$, $[\rho\BV q]_B$ denote the conventional Newton
and Fourier law expressions for a dilute gas ($\rho/\rho_{sp}\ll 1$),
respectively, derived purely from the Boltzmann equation
\cite{Gra,ChaCow}. Observe that the viscosity $\mu$ is $O(\sigma)$.
Therefore, in order to include viscous effects into the fluid dynamics
equations, we will have to retain the $O(\sigma)$-term in the
expansion \eqref{eq:f_expansion}.

\subsection{The Navier--Stokes equations}

Above, we computed the expansion terms $\Mf_0$ and $\Mf_1$ of
\eqref{eq:f_expansion}. The $\Mf_0$ is given via the
Maxwell--Boltzmann state \eqref{eq:MB_state}, while the first-order
correction $\Mf_1$ is given via the Hermite polynomial \eqref{eq:f1},
with the parameters $\BM S$ and $\BV q$ provided via the Newton and
Fourier laws in \eqref{eq:S_q_NS}.

Now, instead of proceeding with the formal hydrodynamic limit as the
sphere diameter $\sigma\to 0$, we will assume that $\sigma$ is a
finitely small constant. In such a case, we can truncate
\eqref{eq:f_expansion} to the first two leading order terms (assuming
that all $O(\sigma^2)$ terms are small enough so that they can be
neglected). The result is known as the Grad state \cite{Gra}:
\begin{equation}
\Mf_{Grad}=\Mf_0+\sigma \Mf_1.
\end{equation}
Then, we follow the same procedure for $\Mf_{Grad}$ as above for
$\Mf_0$; namely, we substitute $\Mf_{Grad}$ into
\eqref{eq:next_order}, and compute the transport equations for the
mass, momentum and energy moments. Since $\Mf_{Grad}$ contains
$O(\sigma)$ terms, the terms of the same order must be retained in the
collision term of \eqref{eq:next_order}. The resulting transport
equation for the density is unchanged, however, the equations for the
velocity and energy are now given via
\begin{subequations}
\label{eq:mom_en_NS_2}
\begin{multline}
\parderiv{(\rho\BV u)}t+\parderiv{}{\BV x}\cdot(\rho(\BV u\BV u^T+
\theta\BM I+\sigma\BM S))=-\frac 4{\rho_{sp}}(1+\sigma R_*)\parderiv{
}{\BV x}\cdot\left(\rho^2\theta\right)+\\+\frac{6\sigma}{\pi\rho_{sp}}
\int(\BV n\cdot(\BV w-\BV v))^2\BV n\Hside\big(\BV n\cdot(\BV w-\BV v)
\big)\\\left(-\Mf_1(\BV v)\BV n\cdot\parderiv{\Mf_0(\BV w)}{\BV x}-
\Mf_0(\BV v)\BV n\cdot\parderiv{\Mf_1(\BV w)}{\BV x}+\frac 12
\Mf_0(\BV v) \BV n^T\parderiv{^2\Mf_0(\BV w)}{\BV x^2}\BV n\right)
\dif\BV n\dif\BV w\dif\BV v,
\end{multline}
\begin{multline}
\parderiv{(\rho\epsilon)}t+\parderiv{}{\BV x}\cdot(\rho((\epsilon
+\theta)\BV u+\sigma\BM S\BV u+\sigma\BV q))=-\frac 4{\rho_{sp}}
(1+\sigma R_*)\parderiv{}{\BV x}\cdot\left(\rho^2\theta\BV u\right)
+\\+\frac{3\sigma}{\pi\rho_{sp}}\int(\BV n\cdot(\BV w+\BV v))(\BV n
\cdot(\BV w-\BV v))^2\Hside\big(\BV n\cdot(\BV w-\BV v)\big)\\\left(
-\Mf_1(\BV v)\BV n\cdot\parderiv{\Mf_0(\BV w)}{\BV x}-\Mf_0(\BV v)\BV
n \cdot\parderiv{\Mf_1(\BV w)}{\BV x}+\frac 12 \Mf_0(\BV v)\BV n^T
\parderiv{^2\Mf_0(\BV w)}{\BV x^2}\BV n\right)\dif\BV n\dif\BV w\dif
\BV v.
\end{multline}
\end{subequations}
Above, for convenience, the constant $R_*$ denotes the derivative of
$R_{\infty,0}^{(2)}$ at zero:
\begin{equation}
R_*=\left.\deriv{}\sigma R_{\infty,0}^{(2)}(\sigma)\right|_{\sigma=0}.
\end{equation}
Due to the fact that both $\Mf_0$ and $\Mf_1$ are given explicitly via
\eqref{eq:MB_state} and \eqref{eq:f1}, respectively, all integrals in
the right-hand side of \eqref{eq:mom_en_NS_2} are computable in terms
of elementary functions. In particular, we arrive at the following
exact relations via direct integration:
\begin{subequations}
\begin{multline}
\int(\BV n\cdot(\BV w-\BV v))^2\BV n\Hside \big( \BV n\cdot(\BV w-\BV
v)\big)\\\left(\Mf_1(\BV v)\BV n\cdot \parderiv{\Mf_0(\BV w)}{\BV x}
+\Mf_0(\BV v)\BV n\cdot\parderiv{\Mf_1(\BV w)}{\BV x}\right)\dif\BV n
\dif\BV w\dif \BV v=\frac{4\pi}{15}\parderiv{}{\BV x}\cdot
\left(\rho^2\BM S\right),
\end{multline}
\begin{multline}
\int(\BV n\cdot(\BV w-\BV v))^2\BV n\Hside\big(\BV n\cdot(\BV w-\BV v)
\big)\BV n^T\parderiv{^2\Mf_0(\BV w)}{\BV x^2}\BV n\Mf_0(\BV v)\dif
\BV n\dif\BV w\dif\BV v=\\=\frac{8\sqrt\pi}{15}\parderiv{}{\BV x}
\cdot\left(\rho^2\sqrt\theta\left(\parderiv{\BV u}{\BV x}+\parderiv{
  \BV u}{ \BV x}^T+\left(\parderiv{}{\BV x}\cdot\BV u\right)\BM I
\right) \right),
\end{multline}
\begin{multline}
\int(\BV n\cdot(\BV w+\BV v))(\BV n\cdot(\BV w-\BV v))^2\Hside \big(
\BV n\cdot(\BV w-\BV v)\big)\\\left(\Mf_1(\BV v)\BV n\cdot\parderiv{
  \Mf_0(\BV w)}{\BV x}+\Mf_0(\BV v)\BV n\cdot\parderiv{\Mf_1(\BV w)
}{\BV x}\right)\dif\BV n\dif\BV w\dif\BV v=\frac{8\pi}{15}\parderiv{
}{\BV x}\cdot\left(\rho^2\left(\BM S\BV u+\frac 32\BV q\right)\right),
\end{multline}
\begin{multline}
\int(\BV n\cdot(\BV w+\BV v))(\BV n\cdot(\BV w-\BV v))^2\Hside\big(\BV
n\cdot(\BV w-\BV v)\big)\BV n^T\parderiv{^2\Mf_0(\BV w)}{\BV x^2}\BV n
\Mf_0(\BV v)\dif\BV n\dif\BV w\dif\BV v=\\=\frac{16\sqrt\pi}{15}
\parderiv{}{\BV x}\cdot\left(\rho^2\sqrt\theta\left[\left(\parderiv{
\BV u}{\BV x}+\parderiv{\BV u}{\BV x}^T+\left(\parderiv{}{\BV x}\cdot
\BV u\right)\BM I\right)\BV u+\frac 52\parderiv\theta{\BV x}\right]
\right).
\end{multline}
\end{subequations}
With the expressions above, the transport equations in
\eqref{eq:mom_en_NS_2} become
\begin{subequations}
\label{eq:mom_en_NS}
\begin{multline}
\parderiv{(\rho\BV u)}t+\parderiv{}{\BV x}\cdot(\rho(\BV u\BV u^T+
\theta\BM I))=-\frac 4{\rho_{sp}}(1+\sigma R_*)\parderiv{}{\BV x}
\cdot\left(\rho^2\theta\right)-\\-\sigma\parderiv{}{\BV x} \cdot\left(
\left(1+\frac{8\rho}{5\rho_{sp}}\right)\rho\BM S\right)
+\frac{8\sigma}{5\sqrt\pi\rho_{sp}}\parderiv{}{\BV x}\cdot\left(
\rho^2\sqrt\theta \left(\parderiv{\BV u}{\BV x}+\parderiv{\BV u}{\BV
  x}^T+\left( \parderiv{}{\BV x}\cdot\BV u\right)\BM I\right)\right),
\end{multline}
\begin{multline}
\parderiv{(\rho\epsilon)}t+\parderiv{}{\BV x}\cdot(\rho(\epsilon
+\theta)\BV u)=-\frac 4{\rho_{sp}}(1+\sigma R_*)\parderiv{}{\BV x}
\cdot\left(\rho^2\theta\BV u\right)-\\-\sigma\parderiv{}{\BV x}
\cdot\left(\left(1+\frac{8\rho}{5\rho_{sp}}\right)\rho\BM S\BV u
\right)-\sigma\parderiv{}{\BV x}\cdot\left(\left(1+\frac{12\rho}{5
  \rho_{sp}}\right)\rho\BV q\right)+\\+\frac{8\sigma}{5\sqrt\pi
  \rho_{sp}} \parderiv{}{\BV x}\cdot\left(\rho^2\sqrt\theta\left[
  \left(\parderiv{ \BV u}{\BV x}+\parderiv{\BV u}{\BV x}^T+\left(
  \parderiv{}{\BV x}\cdot \BV u\right)\BM I\right)\BV u+\frac
  52\parderiv\theta{\BV x}\right] \right).
\end{multline}
\end{subequations}
Substituting the Newton and Fourier laws for $\BM S$ and $\BV q$ from
\eqref{eq:S_q_NS}, we close \eqref{eq:mom_en_NS} under $\rho$, $\BV u$
and $\theta$:
\begin{subequations}
\label{eq:enskog_navier_stokes_R}
\begin{multline}
\parderiv{(\rho\BV u)}t+\parderiv{}{\BV x}\cdot(\rho(\BV u\BV u^T+
\theta\BM I))=-\frac 4{\rho_{sp}}(1+\sigma R_*)\parderiv{(\rho^2
  \theta)}{\BV x}+\\+\parderiv{}{\BV x}\cdot\left(\mu\left((1+a_1)
\left(\parderiv{\BV u}{\BV x}+\parderiv{\BV u}{\BV x}^T\right)-\frac
23(1+a_2)\left(\parderiv{}{\BV x}\cdot\BV u\right)\BM I\right)\right),
\end{multline}
\begin{multline}
\parderiv{(\rho\epsilon)}t+\parderiv{}{\BV x}\cdot(\rho(\epsilon+
\theta)\BV u)=-\frac 4{\rho_{sp}}(1+\sigma R_*)\parderiv{}{\BV x}\cdot
(\rho^2\theta\BV u)+\frac{15}4\parderiv{}{\BV x}\cdot\left(\mu(1+a_3)
\parderiv\theta{\BV x}\right)+\\+\parderiv{}{\BV x}\cdot\left(\mu
\left((1+a_1)\left(\parderiv{\BV u}{\BV x}+ \parderiv{\BV u}{\BV x}^T
\right)-\frac 23(1+a_2)\left(\parderiv{}{\BV x}\cdot\BV u\right)\BM I
\right)\BV u\right).
\end{multline}
\end{subequations}
Above in \eqref{eq:enskog_navier_stokes_R}, the expressions for $a_1$,
$a_2$ and $a_3$ are given via
\begin{subequations}
\begin{equation}
a_1\left(\frac\rho{\rho_{sp}}\right)=\frac{16\rho}{5\rho_{sp}}\left(1+
\frac{4\rho}{5\rho_{sp}} \left(1+\frac{12}\pi\right)\right),
\end{equation}
\begin{equation}
a_2\left(\frac\rho{\rho_{sp}}\right)=\frac{16\rho}{5\rho_{sp}}\left(1+
\frac{4\rho}{5\rho_{sp}}\left(1-\frac{18}\pi\right)\right),
\end{equation}
\begin{equation}
a_3\left(\frac\rho{\rho_{sp}}\right)=\frac{24\rho}{5\rho_{sp}}\left(1+
\frac{2\rho}{15\rho_{sp}} \left(9+\frac{32}\pi\right)\right).
\end{equation}
\end{subequations}
Observe that there are two kinds of $O(\sigma)$-terms in
\eqref{eq:enskog_navier_stokes_R}: the viscous terms with $\mu$, and
the advection terms with $\sigma R_*$. Clearly, the viscous terms
cannot be neglected, since it is common in practice for the velocity
and temperature to have large second derivatives, and thus the
contribution from those terms can be measurable even if $\sigma$ is
small. On the other hand, observe that the advection terms with
$\sigma R_*$ contain only the first derivatives, with the product
$\sigma R_*$ being a constant correction of order $\sigma$ to 1. From
the practical standpoint, the $\sigma R_*$-correction is unlikely to
affect the solution to a measurable extent in most practical
situations, and thus can be discarded in a typical real-world
scenario. The resulting transport equations become
\begin{subequations}
\label{eq:enskog_navier_stokes}
\begin{equation}
 \parderiv\rho t+\parderiv{}{\BV x}\cdot(\rho\BV u)=0,
\end{equation}
\begin{multline}
\parderiv{(\rho\BV u)}t+\parderiv{}{\BV x}\cdot\left(\rho\left(\BV u
\BV u^T+\left(1+\frac{4\rho}{\rho_{sp}}\right)\theta\BM I\right)
\right)=\\=\parderiv{}{\BV x}\cdot\left(\mu\left((1+a_1)\left(
\parderiv{ \BV u}{\BV x}+\parderiv{\BV u}{\BV x}^T\right)-\frac 23
(1+a_2)\left( \parderiv{}{\BV x}\cdot\BV u\right)\BM I\right)\right),
\end{multline}
\begin{multline}
\parderiv{(\rho\epsilon)}t+\parderiv{}{\BV x}\cdot\left(\rho\left(
\epsilon+\left(1+\frac{4\rho}{\rho_{sp}}\right)\theta\right)\BV u
\right)=\frac{15}4\parderiv{}{\BV x}\cdot\left(\mu(1+a_3)\parderiv
\theta{\BV x}\right)+\\+\parderiv{}{\BV x}\cdot\left(\mu\left((1+a_1)
\left(\parderiv{\BV u}{\BV x}+\parderiv{\BV u}{\BV x}^T\right)-\frac
23(1+a_2)\left(\parderiv{}{\BV x}\cdot\BV u\right)\BM I\right)\BV u
\right).
\end{multline}
\end{subequations}
Following the convention set forth in \cite{Lac}, we name the
equations above the ``Enskog--Navier--Stokes'' equations. It is
interesting that similar equations were obtained in \cite{Lac2}, but
the terms which correspond to the Newton and Fourier laws in
\eqref{eq:S_q_NS} were missing. For $\rho/\rho_{sp}\ll 1$ (dilute
gas), the coefficients $a_1, a_2, a_3$ disappear, and the equations in
\eqref{eq:enskog_navier_stokes} become the conventional Navier--Stokes
equations \cite{Bat,Gols}.

Above in \eqref{eq:enskog_navier_stokes}, observe that the presence of
viscous terms is entirely due to the fact that we treat the sphere
diameter $\sigma$ as being finitely small; indeed, if we formally send
$\sigma\to 0$ as in the constant-density hydrodynamic limit above, the
viscous terms in \eqref{eq:enskog_navier_stokes} will disappear, and
the Enskog--Euler equations \eqref{eq:enskog_euler} will emerge
instead.

\section*{Summary}

In the current work, we start by examining the conventional derivation
of the Boltzmann equation \cite{Bol,Gols,Gra} from the Liouville
equation for the hard sphere dynamics \cite{Cer2,CerIllPul,GalRayTex}.
The conventional derivation consists, roughly, of two parts -- the
first part is the derivation of the BBGKY hierarchy
\cite{Bog,BorGre,Kir} from the Liouville equation under the assumption
that the solution of the Liouville equation is known, while the second
part is the derivation of the Boltzmann equation from the leading
order BBGKY identity by, first, altering the integrand under the
collision integral (Boltzmann hierarchy), and second, factorizing the
joint probability distribution of two spheres into the product of the
single-sphere distributions. We observe that the second part of the
derivation contradicts the assumptions under which the first part of
the derivation was carried out. We also note that some of the observed
contradictions are similar to the objection pointed out by Loschmidt
\cite{Los}.

At the same time, it is known from observations and experiments that
the Boltzmann equation is {\em de facto} an excellent model for a
dilute gas in practical scenarios. We, therefore, propose that there
should exist a different underlying dynamical model of the hard sphere
gas, which does not violate Loschmidt's objection and, at the same
time, leads to the Boltzmann equation in the dilute gas approximation.
We subsequently formulate a random process with an infinitesimal
generator, which triggers the necessary velocity jumps via a suitable
point process \cite{DalVer,Papa} whenever the collision condition is
detected. This process is a L\'evy-type Feller process
\cite{App,Fel2}, and the general form of its characteristic function
is given by Courr\`ege's theorem \cite{Cou}. We subsequently formulate
the forward Kolmogorov equation \cite{GikSko} for the probability
density of the random dynamics, compute those of its steady states
which are uniform in the velocity variables, and show that the
corresponding Kullback--Leibler entropy \cite{KulLei} of the complete
system of spheres is a nonincreasing function of time.

We find that, in the case of many spheres, which are distributed
independently and identically at the initial time, the corresponding
steady state is uniform not only in velocities, but also in the
coordinates of the spheres, except for the ``contact zones'' (that is,
the sets of coordinates which satisfy the collision condition). With
help of the computed marginal distributions of these steady states, we
derive the forward equation for the dynamics of a single sphere under
the assumption that the distribution of the full system is invariant
under the reordering of the spheres.

As the total number of spheres in the system becomes large, we find
that, for the limiting dynamics of thin contact zones and impenetrable
spheres, the forward equation becomes a variant of the Enskog equation
\cite{BelLac,Ens,GapGer,Lac,vBeiErn}. Further, assuming that the gas
is dilute, we arrive at the Boltzmann equation
\cite{Bol,Cer,Cer2,CerIllPul,Gols,Gra}. Finally, we find that, in the
hydrodynamic limit of constant-density spheres, the resulting
Enskog--Euler and Enskog--Navier--Stokes equations acquire additional
effects in both the advective and viscous terms, which are absent in
the conventional Euler and Navier--Stokes equations derived from the
Boltzmann equation.

\ack The author thanks Prof. Ibrahim Fatkullin and Prof. Peter Kramer
for helpful discussions. The work was supported by the Office of Naval
Research grant N00014-15-1-2036.

\end{document}